\documentclass[sigconf, authorversion]{acmart}

\AtBeginDocument{%
  \providecommand\BibTeX{{%
    \normalfont B\kern-0.5em{\scshape i\kern-0.25em b}\kern-0.8em\TeX}}}

\setcopyright{acmlicensed}

\copyrightyear{2026}
\acmYear{2026}
\setcopyright{cc}
\setcctype{by}
\acmConference[CHI '26]{Proceedings of the 2026 CHI Conference on Human Factors in Computing Systems}{April 13--17, 2026}{Barcelona, Spain}
\acmBooktitle{Proceedings of the 2026 CHI Conference on Human Factors in Computing Systems (CHI '26), April 13--17, 2026, Barcelona, Spain}
\acmPrice{}
\acmDOI{10.1145/3772318.3790453}
\acmISBN{979-8-4007-2278-3/2026/04}

\usepackage{enumitem}
\setlist[itemize]{leftmargin=*}
\setlist[enumerate]{leftmargin=*,label=\arabic*.}

\usepackage[shortcuts]{extdash}

\usepackage{soul}

\usepackage{xspace}

\usepackage{xparse}

\newlength{\cpsp}

\newcommand{\qt}[1]{\textit{``#1''}}
\newcommand{\pqt}[2]{\textit{``#1''}{\,}{\small-#2}}

\newcommand{\ie}{i.e.\@\xspace}
\newcommand{\eg}{e.g.\@\xspace}

\newcommand{\etal}{et~al.\@\xspace}

\NewDocumentCommand\p{ m g }{%
    \IfNoValueTF{#2}{%
         {$p<#1$}%
    }{%
         {$p#2#1$}%
    }%
}

\NewDocumentCommand\anova{ m m m m g }{%
    \IfNoValueTF{#5}{%
         {$F_{#1,#2}=#3$, $p<#4$}%
    }{%
         {$F_{#1,#2}=#3$, $p<#4$, $\eta_G^2=#5$}%
    }%
}

\PassOptionsToPackage{dvipsnames}{xcolor}

\newenvironment{commentwrapper}[1]{\color{#1}}{\color{black}}

\newcommand{\authorcomment}[3]{\begin{commentwrapper}{#1}[#2:~#3]\end{commentwrapper}}

\newcommand{\simplecomment}[2]{\begin{commentwrapper}{#1}#2\end{commentwrapper}}

\definecolor{GREEN}{rgb}{0.0,0.6,0.0}
\definecolor{BLUE}{rgb}{0.0,0.2,0.7}
\definecolor{GOLD}{rgb}{0.6,0.6,0.0}
\definecolor{CYAN}{rgb}{0.0,0.5,0.5}
\definecolor{PURPLE}{rgb}{0.5,0.0,0.5}

\definecolor{RED}{rgb}{0.7,0.0,0.0}
\definecolor{GRAY}{gray}{0.5}

\definecolor{LIGHTGRAY}{HTML}{DDDDDD}

\newcommand{\outline}[1]{
\medbreak
\noindent
{\sethlcolor{LIGHTGRAY}\hl{> #1}}
\medbreak
}

\definecolor{GOLD}{HTML}{FFE052}
\newcommand{\revnote}[1]{
\medbreak
\noindent
{\sethlcolor{GOLD}\hl{> #1}}
\medbreak
}

\newcommand{\fixme}[1]{\simplecomment{RED}{#1}}

\newcommand{\hideoutline}{
    \renewcommand{\outline}[1]{} 
}

\newcommand{\hiderevnotes}[0]{
    \renewcommand{\revnote}[1]{} 
}

\newcommand{\hidecomments}[0]{
    \hideoutline
    \hiderevnotes
    \renewcommand{\authorcomment}[3]{} 
    \renewcommand{\fixme}[1]{\simplecomment{black}{##1}}
}

\newcommand{\rev}[1]{#1}

\definecolor{REVISIONRED}{HTML}{DE3163}
\definecolor{REVISIONGREEN}{HTML}{9FE2BF}
\definecolor{REVISIONYELLOW}{HTML}{FFBF00}
\definecolor{REVISIONBLUE}{HTML}{6495ED}

\PassOptionsToPackage{bookmarks}{hyperref}

\graphicspath{{figs/}{figures/}{pictures/}{images/}{./}} 

\hidecomments

\usepackage{fancybox,xcolor} %
\newcommand{\btn}[1]{%
  \begingroup
  \setlength{\fboxsep}{1pt}%
  \ovalbox{%
    \colorbox{black!10}{%
      \sffamily\raisebox{0pt}[1.4ex][0.3ex]{#1}%
    }%
  }%
  \endgroup
}

\newcommand{\peek}{
    \raisebox{-0.5ex}
    {\includegraphics[height=1em]{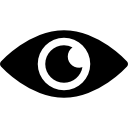}}
}

\newcommand{\slide}{
    \raisebox{-0.4ex}
    {\includegraphics[height=1em]{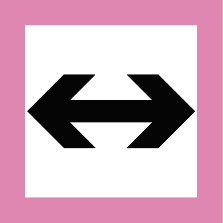}}
}

\newcommand{\extend}{
    \raisebox{-0.4ex}
    {\includegraphics[height=1em]{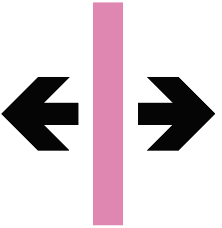}}
}

\newcommand{\swap}{
    \raisebox{-0.4ex}
    {\includegraphics[height=0.8em]{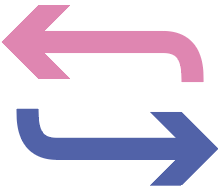}}
}

\begin{document}

\title{To Slide or Not to Slide: Exploring Techniques for Comparing Immersive Videos
}

\author{Xizi Wang}
\affiliation{%
  \institution{University of Waterloo}
  \city{Waterloo}
  \country{Canada}}
\email{l84wang@uwaterloo.ca}

\author{Yue Lyu}
\affiliation{%
  \institution{University of Waterloo}
  \city{Waterloo}
  \country{Canada}}
\email{yue.lyu@uwaterloo.ca}

\author{Yalong Yang}
\affiliation{%
  \institution{Georgia Institute of Technology}
  \city{Atlanta}
  \country{USA}}
\email{yalong.yang@gatech.edu}

\author{Jian Zhao}
\affiliation{%
  \institution{University of Waterloo}
  \city{Waterloo}
  \country{Canada}}
\email{jianzhao@uwaterloo.ca}

\renewcommand{\shortauthors}{Wang et al.}

\begin{teaserfigure}
    \includegraphics[width=\textwidth]{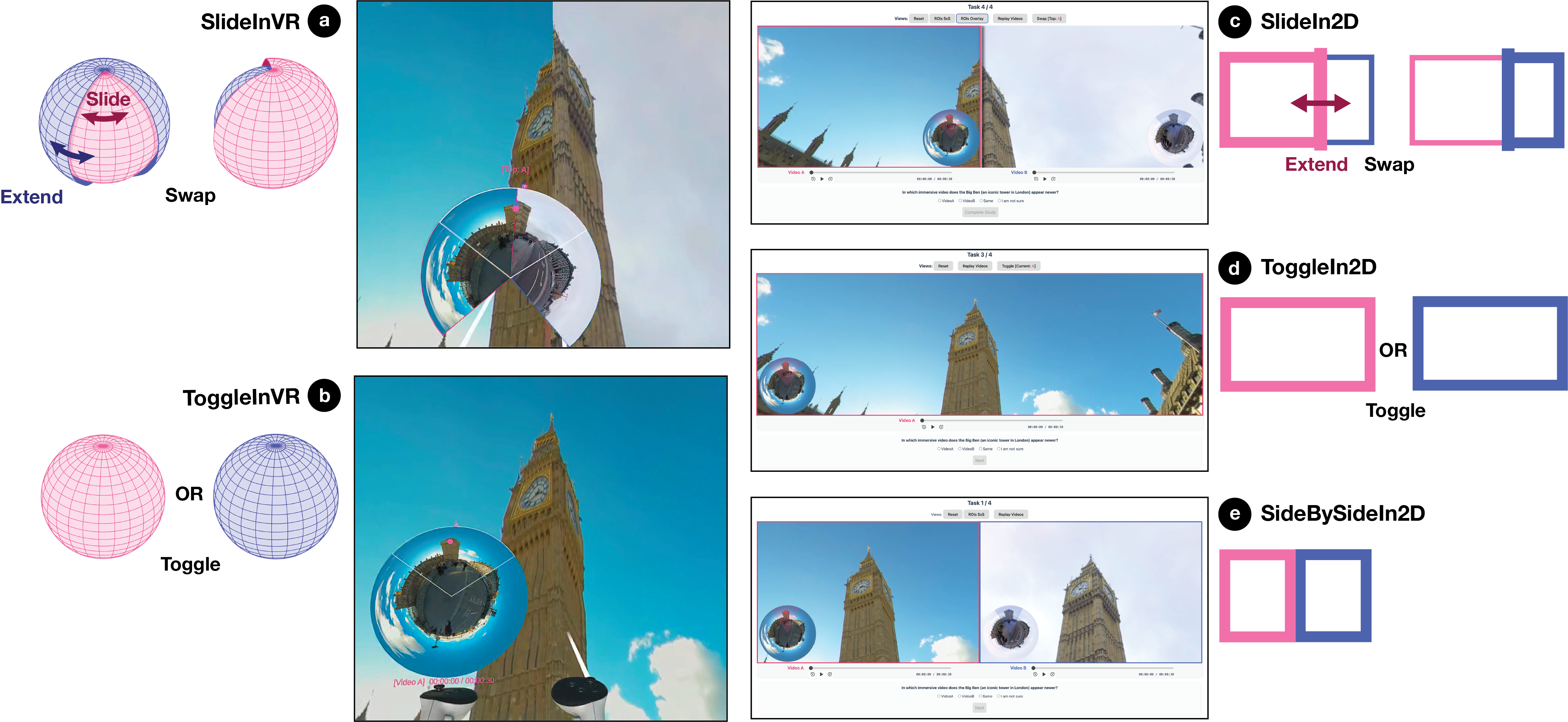}
    \caption{Overview of the five techniques for comparing two immersive 360\textdegree{} videos, each illustrated with a schematic of key interactions and a task example: (a) SlideInVR, (b) ToggleInVR, (c) SlideIn2D, (d) ToggleIn2D, and (e) SideBySideIn2D.
    Techniques (a) and (b) support in-headset VR comparison, and techniques (c)–(e) are presented in 2D desktop settings. 
    All examples depict the same \rev{Task\#14, where two immersive videos from © VR Gorilla (used with permission) captured the same Big Ben from two nearby camera viewpoints in 2017~\cite{VR_Gorilla_old_london} and 2025~\cite{VR_Gorilla_new_london}.}
    }
    \label{fig:teaser}
    \Description{Overview of the five techniques for comparing two immersive 360-degree videos, each illustrated with a schematic of key interactions and a task example: (a) SlideInVR, (b) ToggleInVR, (c) SlideIn2D, (d) ToggleIn2D, and (e) SideBySideIn2D. 
    Techniques (a) and (b) support in-headset VR comparison, and techniques (c)–(e) are presented in 2D desktop settings. 
    All examples depict the same Task\#14, where two immersive videos (© VR Gorilla; used with permission) captured the same Big Ben from two nearby camera viewpoints in 2017 and 2025.}
\end{teaserfigure}

\begin{abstract}
Immersive videos (IVs) provide 360\textdegree{} environments that create a strong sense of presence and spatial exploration. 
Unlike traditional videos, IVs distribute information across multiple directions, making comparison cognitively demanding and highly dependent on interaction techniques.
With the growing adoption of IVs, effective comparison techniques have become an essential yet underexplored area of research.
Inspired by the ``sliding'' concept in 2D media comparison, we integrate two established comparison strategies from the literature---toggle and side-by-side---to support IV comparison with greater flexibility.
For an in-depth understanding of different strategies, we \rev{adapt and implement} five IV comparison techniques across VR and 2D environments: SlideInVR, ToggleInVR, SlideIn2D, ToggleIn2D, and SideBySideIn2D.
We then conduct a user study ($N=20$) to examine how these techniques shape users' perceptions, strategies, and workflows. 
Our findings provide empirical insights into the strengths and limitations of each technique, underscoring the need to switch between comparison approaches across scenarios.
Notably, participants consistently rate SlideInVR and SlideIn2D as the most flexible and favorite methods for IV comparison.

\end{abstract}



\maketitle

\section{Introduction}
Immersive video (IV), \rev{\eg, 360\textdegree{} video,} has gained significant attention due to its ability to provide a more engaging, interactive, and realistic experience compared to traditional 2D videos, fostering greater focus, interest and active participation as well as promoting effective and low-workload learning ~\cite{IV_benefits_high_edu:2022, immersivePOV:CHI22, IV_NursingEdu:2024}.
It is widely used across various domains, including general education (\eg, used as lecture materials or lecture recording~\cite{VR_RealPedagogy:CHI24,IV_benefits_high_edu:2022,IV_Education_ReviewPaper:2023,Aukland_360_video_as_ethnographic_material, climate_change_comparison:Markowitz:2018}), professional training (\eg, nursing~\cite{IV_NursingEdu:2024} and surgeon training~\cite{IV_Surgery:2021}), and entertainment (\eg, sports or concert recordings and narrative video content~\cite{IV_Characterization:2017, VR_Gorilla_new_london}). 
\rev{As IV adoption grows, practitioners (and sometimes viewers) may frequently need to compare two IV clips during production (\eg, reviewing and selecting IV clips~\cite{Freeman_video_workflow} across takes or viewpoints~\cite{JumpCut_CinematicVR_30and180-degree:2024, Murch:Blink:2021}), editing (\eg, comparing stitched IV outputs produced by different stitching tools~\cite{Altman_2017_editing_360_video} and aligning the points of interest between two clips~\cite{Nguyen:Vremiere:CHI17}), and evaluation (\eg, making pairwise quality selections within a larger IV collection~\cite{ITU-T-P919-2020}, evaluating the quality and consistency of AI-generated IVs across AI models~\cite{360DVD:2024}, and comparing the visual or motion differences of two IV recordings during learning activities~\cite{Aukland_360_video_as_ethnographic_material}).}
\rev{However, 2D video comparison workflows do not directly translate to IV comparison, because viewers need to navigate a omnidirectional scene where maintaining ego-centric spatial context is paramount; without specialized techniques, this process often leads to significant memory strain and disorientation.}

Virtual Reality (VR) is the most common modality for viewing and comparing 360\textdegree{} IVs, %
offering a direct and intuitive visual representation of the IV content.
This approach allows users to fully experience the video and make effective comparisons and decisions.
However, VR typically supports only one IV at a time, forcing users to toggle between videos during comparison. 
This disrupts workflows, increases reliance on short-term memory to retain visual details between toggles, and ultimately reduces both the accuracy and efficiency of IV comparisons~\cite{Gleicher:VisualComparison_taxonomy:InFoVis:2011}.
While prior research has explored the navigation and editing of IVs in VR~\cite{Nguyen:Vremiere:CHI17,Griffin:6DIVE:2021}, these efforts largely focus on single-video interaction rather than supporting direct comparison \rev{of different} IVs.

In contrast, projecting IVs onto a 2D screen treats them as traditional flat videos but sacrifices essential spatial information. 
Flattening spherical content introduces distortions, which can obscure important details and hinder accurate comparisons ~\cite{Nguyen:Vremiere:CHI17}. 
Previous work on 2D video comparison has explored approaches such as side-by-side, toggle/overlay, and explicit encoding ~\cite{Baker:ComparingVideo:GI24,VisualComparator:2020}, demonstrating their efficiency for traditional videos.
However, their applicability to immersive 360\textdegree{} content remains underexplored, leaving open questions about how to effectively support IV comparisons.

These limitations in both VR and 2D highlight the lack of a general approach that supports direct, flexible, and efficient comparison of 360\textdegree{} IVs.
To fill in the gap, we investigate how different comparative approaches (\eg, toggle, side-by-side) and modalities (\eg, VR, 2D) affect users when comparing two IVs. 
Inspired by the techniques in 2D video comparison \cite{Baker:ComparingVideo:GI24,VisualComparator:2020}, we adapt the general concept of ``sliding''\rev{~\cite{Baker:ComparingVideo:GI24,KnightLabJuxtapose,Tominski_Forsell_Johansson_Natural_Behavior_2012}, which is commonly used for comparing 2D images and videos, to support pairwise comparison of IVs in both 2D and VR.}
This adapted sliding concept lets users dynamically transition between side-by-side and toggle views, effectively combining them into a hybrid approach.
Concretely, it overlays two IVs as front and back layers and comprises three core features (\textit{slide}, \textit{extend}, and \textit{swap}) that support flexible adjustments of positions and visible regions of IVs.
\textit{Slide} allows users to reposition the two IVs to place the regions of interest (ROIs) side by side or in an overlapping (toggle-like) configuration;
\textit{extend} enables users to adjust the spatial distribution by extending or shrinking the visible area of each IV within a limited view;
and \textit{swap} switches the order of the two layers, changing which IV is prioritized and visible.

We apply this concept to VR and 2D modalities, resulting in two techniques: SlideInVR (\autoref{fig:teaser}-a) and SlideIn2D (\autoref{fig:teaser}-c). 
\rev{Employing just the toggle and side-by-side approaches}, we further \rev{implement} three basic IV comparison techniques, including ToggleInVR (\autoref{fig:teaser}-b), ToggleIn2D (\autoref{fig:teaser}-d) and SideBySideIn2D (\autoref{fig:teaser}-e). 
There is no SideBySideInVR because placing two omnidirectional IVs directly side by side in VR is not feasible.
To better support IV comparisons, all techniques provide zoomed-in and global views as well as temporal-spatial visualizations of ROIs.

To address our overarching goal of exploring how comparative approach and modality can affect user' perceptions, strategies, and workflows, we conducted a within-subject study with 20 VR users to assess the pros and cons of these five techniques. 
We derive four types of IV comparison tasks, including comparisons of temporal occurrence, spatial-temporal motions, local visual differences, and global visual differences, from previous work~\cite{Amar:Low-Level_Components_of_Analytic_Activity_in_Information_Visualization:2025,Baker:ComparingVideo:GI24}.
Our results highlighted the benefits and trade-offs of different comparison approaches and modalities.
VR was perceived to offer more natural navigation, spatial awareness, and effectiveness in spotting visual quality and distortions, but also introduced a higher workload and a steeper learning curve compared to 2D.
Both task type and technique shaped participants’ workflows and strategies: 
side-by-side comparison was generally preferred for temporal occurrence and dynamic ROIs, while toggle was more effective for fine-grained visual differences when ROIs appeared in the same location.
Notably, SlideInVR and SlideIn2D consistently emerged as the two most favored techniques, valued for their flexibility and ability to hybrid multiple comparison methods while supporting dynamic allocation of space.

\rev{In summary, our work makes the following contributions: (1) we extend the concept of sliding in 2D image/video comparison to immersive video comparison, resulting in two new techniques, SlideInVR and SlideIn2D, which enable seamless transitions between toggle and side-by-side approaches;
(2) we conduct a within-subjects study (N=20) evaluating five representative immersive video comparison techniques across 2D and VR as well as comparison approaches, which provide empirical findings and design guidance on how modality and task characteristics shape users’ workflows and strategies.}

\section{Background}
\subsection{Immersive Videos and Applications}
Immersive videos (IVs) refer to video formats that capture a scene in a wide field of view (FoV), most commonly including 180\textdegree{} and 360\textdegree{} videos (also referred to as omnidirectional, panoramic, or spherical videos~\cite{A_Survey_on_360_Video_Streaming:Fan:2019}). 
Several common 2D projection formats are used to represent IVs.
The equirectangular projection, which provides a panoramic overview, is the most common projection used by major platforms such as YouTube~\cite{A_survey_on_360-degree:2021:Federico}.
The azimuthal equidistant projection, rendered as a ``planet view'', has been applied for visualization and minimap applications, but is less common as a primary distribution format~\cite{A_survey_on_360-degree:2021:Federico, VR_planet:Fan:2016, Nguyen:Vremiere:CHI17}.
While these projections make spherical video compatible with conventional 2D players, they also introduce geometric distortion, which can affect perceived visual quality and usability~\cite{Nguyen:Vremiere:CHI17,A_survey_on_360-degree:2021:Federico}.
IVs can be captured and displayed in either monoscopic or stereoscopic modes~\cite{evaluating_user_experience_of_180_and_360_degree_images:2020}. 
The monoscopic is one of the most common formats and supported by most consumer devices; while stereoscopic provides a stronger sense of depth perception, it requires more complex capture. 
Therefore, in this paper, our focus is comparing monoscopic 360\textdegree~IVs.

IVs have been widely applied in various domains, including higher education~\cite{IV_Education_ReviewPaper:2023, IV_benefits_high_edu:2022, VR_RealPedagogy:CHI24,Aukland_360_video_as_ethnographic_material}, virtual field trips~\cite{climate_change_comparison:Markowitz:2018,historical_tour:Abidin:2020}, professional training~\cite{IV_NursingEdu:2024}, and entrainment~\cite{IV_Characterization:2017}.
When viewed through VR devices, IVs provide immersive environments that foster a strong sense of presence and spatial exploration~\cite{IV_benefits_high_edu:2022}. 
Compared to traditional 2D videos, they also enable viewers, especially learners, to feel more focused, motivated, and joyful, encouraging active engaging and interactions~\cite{IV_benefits_high_edu:2022, IV_NursingEdu:2024, IV_Surgery:2021}.
These benefits and applications highlight the need for better support in comparing different types of IVs and understanding their differences.
For example, Jin \etal found that instructors expressed their concerns about the quantity and quality of IV when selecting and producing educational materials~\cite{VR_RealPedagogy:CHI24}.
Motivated by these opportunities and challenges, our work explores how different modalities can support experts and general users in comparing different types of IVs.

\subsection{Immersive Videos Editing}
There exist several commercial applications, along with powerful plugins, for IV editing, such as Insta360 Studio\footnote{https://www.insta360.com/download} and Adobe Premiere Pro\footnote{https://www.adobe.com}. 
For example, Mettle’s Skybox VR Player allowed users to preview spherical video within Premiere, while Adobe later showcased CloverVR, an in-VR editing extension for Premiere, at Adobe MAX 2016~\cite{Clover_2016}.
The research community has also explored various tools and techniques to support IV editing.
For instance, Nguyen \etal proposed Vremiere that allows users to browse and edit monoscopic IVs directly in a VR headset while still relying on mouse and keyboard as input devices~\cite{Nguyen:Vremiere:CHI17}. 
This in-headset interface introduced a new VR editing workflow and demonstrated several helpful features, including the overview of IVs in planet view that we later adopted in our design.
They also highlighted the importance of WYSIWYG (``What You See Is What You Get'') during editing for reducing the gulf between traditional 2D desktop editing and immersive VR viewing.
Building on this work, 6DIVE is a system supporting fully in-VR stereoscopic IV editing and multi-track spatial editing that enables users to blend one video clip on top of another similar to picture-in-picture~\cite{Griffin:6DIVE:2021}. 
A comparison of 6DIVE with a desktop editor showed that VR allowed more intuitive spatial alignment and received higher attractiveness and stimulation.
\rev{While these systems focus on IV editing and preview, they highlight recurring moments where editors compare alternative clips or edit variants (\eg, clip selection, alignment, and placement)~\cite{Freeman_video_workflow, Griffin:6DIVE:2021, Nguyen:Vremiere:CHI17} and imply a need for comparison techniques that work directly in VR, motivating a more systematic study of IV comparison techniques.
}

\subsection{Navigation in Immersive Videos}
Researchers have also explored how to improve the navigation in immersive videos, which is a fundamental task in IV editing, browsing, and other applications. 
The first body of research has focused on physical navigation in IVs, which is how viewers use natural body movements, such as head rotations, leaning, or walking, to explore IV content.
For example, shot orientation techniques were proposed to automatically reorient the viewpoint during IV viewing to keep important content in view~\cite{shot_Orientation_Controls:UIST:2017}.
Similarly, Lin \etal investigated automatic rotation and visual guidance techniques, showing that they improved users’ ability to maintain focus across different types of IVs~\cite{Tell_Me_Where_to_Look:CHI_2017}.
Kang and Cho introduced an interactive navigation system for IV playback that automatically generates a viewing path based on saliency and optical flow, allowing users to adjust the view interactively to improve comfort and reduce missed content~\cite{automatic_navigation_for_360:Kang:2019}.

Some works have also explored alternative projections and additional views of IVs to support navigation and reduce the fear of missing out.
For example, Li \etal presented an algorithm for generating route tapestries (a new projection showing left and right side views) to support navigating tour-specific IVs and comparing vacation destinations in a 2D interface~\cite{Route_Tapestries:Li:2021:UIST}.
Vermast and Hürst proposed two 3D thumbnail designs for providing overviews of IVs, and found that spherical 3D thumbnails improved user experience and supported information searching in VR~\cite{3D_Thumbnails_360:Vermast:2023}.
Lin \etal presented Outside-In, a technique that visualizes out-of-sight ROIs in IV by reintroducing them as spatial picture-in-picture previews~\cite{Outside-In:Lin:2017}. 
Prior studies also explored the use of panoramic thumbnails and highlighted the importance of providing an overview during IV viewing~\cite{Panoramic_Thumbnail_fear_of_missing:2021}.

Besides physical navigation, other work has focused on temporal navigation, which is how viewers control their position in time within an IV, such as rewinding, fast-forwarding, or jumping to specific moments.
Yu and Bi studied different types of IV timelines in VR and found that 3D wrist-based timeline offered better user experiences~\cite{3D_timeline:Yu:2023}.
Also, Liu \etal presented RadarVR that provides spatiotemporal visual guidance for ROIs to support IV viewing~\cite{RadarVR:2023:Liu:UIST}.
These findings inspired our design of temporal navigation around the controller and encouraged us to further explore more natural forms of temporal navigation in IVs.

Together, these systems demonstrate a variety of approaches for physical and temporal navigation. However, as another critical task, little work has explored IV comparison across 2D or VR modalities. 
To address this gap, we examine five IV comparison techniques across 2D and VR, revealing how these techniques, task type and modality influence users’ navigation, strategies, and overall perceptions.

\subsection{Visual Comparison}
Prior work in visualization research has established taxonomies of comparative designs.
Gleicher \etal summarized a taxonomy of comparative designs and identified three main comparison techniques: juxtaposition (side-by-side), superposition (overlay), and explicit encoding~\cite{Gleicher:VisualComparison_taxonomy:InFoVis:2011}.
In juxtaposition, items are shown in separate views for direct comparison, which we applied directly in our design of SideBySideIn2D.
Superposition overlays items in the same space and often relies on techniques such as semi-transparency or quick toggles to animate between views for easier comparison. 
We adopted this approach in our ToggleIn2D and ToggleInVR techniques.
Explicit encoding visualizes the relationships or differences between these two items directly.
These categories provide a foundation for situating subsequent work on visual comparison in both 2D and immersive environments, regardless of the data format.

\subsubsection{Visual Comparison in 2D}
Visual comparison has been widely explored in 2D visualization.
For example, Tominski, Forsell, and Johansson investigated a natural folding interaction alongside other comparative interactions (\ie, side-by-side and shine-through) for comparing printed complex data on a 2D interface~\cite{Tominski_Forsell_Johansson_Natural_Behavior_2012}.
Our ``sliding'' concept was initially inspired by the natural interaction with ``paper'' demonstrated in this work, which provided high flexibility for switching between different comparison approaches by moving papers around.

Beyond this work, other studies also explored different comparison approaches in 2D interfaces, especially traditional 2D video comparison.
Cherukuru and Scheitlin demonstrated a visual comparator that supports juxtaposition, superposition, and explicit encoding for comparing dynamic spatiotemporal data.
Their system allows users to superimpose up to three synchronized animations and compare them through a slider-based interface, helping reduce visual clutter while drawing attention to temporal changes.
Furthermore, these ``before and after'' slider widgets have been widely adopted in commercial 2D image comparison tools and web platforms (\eg, Juxtapose~\cite{KnightLabJuxtapose}), where users drag a divider to reveal more or less of each image.
Our IV comparison techniques aligned with this design rationale, particularly the SlideIn2D where juxtaposition, superposition, and sliding interactions are adapted to IVs to support both overview and subtle difference detection.
Recently, Baker \etal introduced a suite of interaction techniques for 2D video comparison~\cite{Baker:ComparingVideo:GI24}. 
Their system supports equalizing the reference frames of videos, juxtaposition techniques such as side-by-side and small multiples, superposition techniques including transparency and window-blind overlays, explicit encodings of differences, and timeline-based translations between temporal and linear representations. 
These techniques reduce the manual effort required for comparing 2D videos and improve usability across diverse datasets. 
Our IV comparison techniques are closely related in that we also adapt juxtaposition and superposition comparison.
However, we extend this design space to immersive 360\textdegree~video formats and emphasize flexible switching between comparison approaches. 
Unlike 2D videos in their study, which often share a fixed perspective, IV introduces additional challenges, such as the need to process information from all directions, viewing multiple non-distorted 360\textdegree~videos simultaneously, tracking moving objects in space, balancing local and global comparisons, and comparing across different viewpoints, which our work specifically explored.

\subsubsection{Visual Comparison in VR}
Visual comparison in VR has received growing attention, as immersive environments introduce unique challenges and opportunities beyond traditional 2D settings.
Recently, Yuan \etal explored different VR techniques for comparing color differences in 360\textdegree~images, including a side-by-side layout that placed 2D images next to each other in VR, a sequential approach that allowed users to switch between full images, and a partial approach that displayed portions of each 360\textdegree~image~\cite{360_image_comparison:Yuan:2025}. 
While their design rationale aligns with ours, their techniques focus primarily on detecting subtle differences of static images in VR and offer less flexibility than our IV comparison approaches.
Another recent work by Zhu \etal evaluated three 3D visual comparison techniques for change detection in VR~\cite{3D_visual_comparison:Zhu:2025}. 
These techniques, including Sliding Window, 3D Slider and Switch Back, were derived from the classic paradigms of juxtaposition and superposition comparison.
Their design concept of toggle-style switching (Switch Back) and slider-based partitioning (Sliding Window and 3D Slider) also parallels our IV comparison techniques (toggle and sliding) and features (peeking). 
However, whereas Zhu \etal focused on 3D scene change detection, our work explores how different comparison approaches can be flexibly adapted for IV comparison tasks.

In motor-skill learning, Ma \etal introduced avaTTAR, an AR system that combines detached and on-body visualizations to support table tennis stroke learning~\cite{Ma:avaTTAR:UIST:2024}. 
Their system juxtaposes expert and user avatars for detached (side-by-side) comparison while also overlaying on-body cues to guide posture correction. 
This highlights suitable usage cases for different comparison approaches: side-by-side supports overall comparisons, whereas overlay cues help users perceive local subtle differences, consistent with findings from prior work~\cite{Gesture_Guidacne:Wang:2024}.
While their focus is on motor-skill acquisition, our work investigates immersive video comparison techniques that similarly build on juxtaposition and superposition, but emphasize flexible modalities for analyzing visual differences across 360\textdegree~video content.

\section{The Sliding Concept}

\begin{figure}[tb]
    \includegraphics[width=0.47\textwidth]{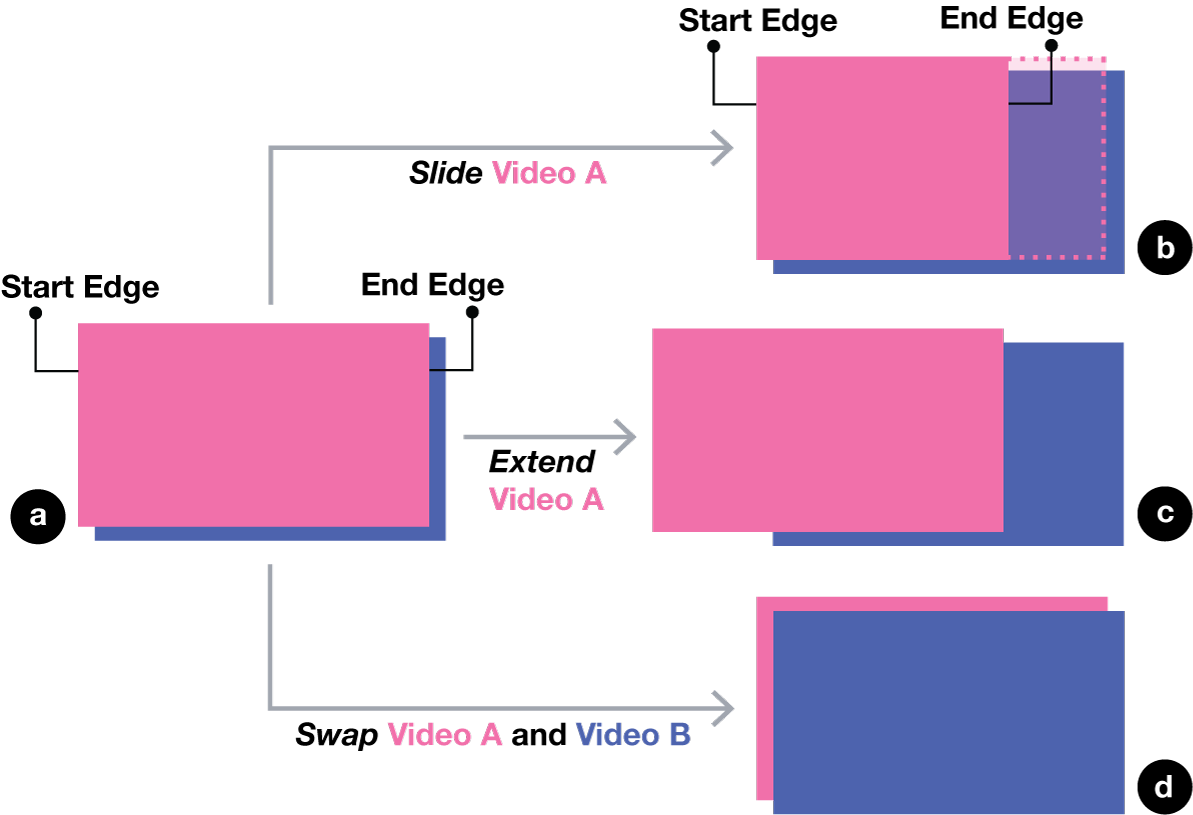}
    \vspace{-2mm}
    \caption{
Abstract illustration of the three core features of our adapted sliding concept in panels (a)--(d). 
Two IVs, Video~A (pink) and Video~B (blue), are shown in overlapping display areas in (a). 
The concept supports three interactions: \textit{extend}\extend, from (a) to (b); \textit{slide}\slide, from (a) to (c); and \textit{swap}\swap, from (a) where the front layer is Video~A to (d) where the front layer is Video~B.
}
    \centering
    \label{fig:sliding}
    \Description{
Abstract illustration of the three core features of our adapted sliding concept in panels (a)--(d). Two immersive videos, Video A (pink) and Video B (blue), are shown on overlapping display areas in (a). The concept supports three interactions: extend, from (a) to (b); slide, from (a) to (c); and swap, from (a) where the front layer is Video A to (d) where the front layer is Video B.
}
    \vspace{-4mm}
\end{figure}

Prior work on 2D comparison systems has demonstrated a variety of approaches, including toggle, side-by-side, and explicit encoding~\cite{Baker:ComparingVideo:GI24,visual_comparison_nature:Tominski:2012,Gleicher:VisualComparison_taxonomy:InFoVis:2011}. 
When applied to 360\textdegree{} videos, each approach has its own limitations.
Toggle enables users to overlay two IVs and switch between them temporally. 
This preserves spatial alignment and supports fine-grained comparison when the ROIs align, but meanwhile places a heavy burden on short-term visual memory~\cite{Gleicher:VisualComparison_taxonomy:InFoVis:2011}, especially for long IVs. 
This challenge is amplified in VR: if the ROIs are not in the same viewing direction, users must rotate their viewpoint between toggles, further increasing memory load and also physical demanding.
Side-by-side, in contrast, allows holistic comparisons by showing both contexts simultaneously in 2D.
However, splitting the FoV into separate views disrupts spatial continuity and forces users to mentally align disparate perspectives~\cite{Gleicher:VisualComparison_taxonomy:InFoVis:2011}. 
In VR, a true side-by-side presentation is not feasible for 360\textdegree{} IVs: a viewer can only be immersed inside one omnidirectional sphere at a time, so we cannot simply place two full 360\textdegree{} environments ``next to each other.'' 
Instead, side-by-side designs for 360\textdegree{} IVs necessarily present each IV as a partial view (\eg, in windows or cropped regions), which shrinks each view and limits peripheral context and immersion.
Additionally, explicit encoding is effective when relationships can be computed and visualized, especially in data visualizations~\cite{Gleicher:VisualComparison_taxonomy:InFoVis:2011}.
But in texture-rich 360\textdegree{} video, such relationships are difficult to define or calculate.
Thus, in this work, we are not considering explicit encoding for 360\textdegree{} videos comparison.
These limitations motivate our ``sliding'' approach for IV comparison, which combines the strengths of toggle and side-by-side for IV comparison tasks. %

While sliding is not a new interaction in 2D media comparison, we significantly extend and adapt this concept to IV comparison, particularly in VR.
In sliding, each IV is represented as a display area defined by a start and an end edge (\autoref{fig:sliding}-a), and these display areas are layered in terms of rendering. 
Users can configure the static layout of the display areas by adjusting their sizes and positions relative to others.
Sliding supports three core interactions (\autoref{fig:sliding}-a,b,c,d):
\begin{enumerate}
    \item \textit{extend}\extend: extending the edges of each IV display area to control how much of the 360\textdegree{} field is revealed;
    \item \textit{slide}\slide: repositioning the display area around the polar axis of the spherical video to adjust the layout;
	\item \textit{swap}\swap: changing the layer order to foreground a different IV.
\end{enumerate}
Together, these interactions enable users to flexibly transition between overlay and side-by-side views, supporting both fine-grained and holistic comparisons of media.
While the sliding concept can in principle support comparison of multiple IVs, in this work, we focus on two-IV comparison and implement features specifically for this case.

\begin{figure*}[htb]
    \includegraphics[width=0.64\textwidth]{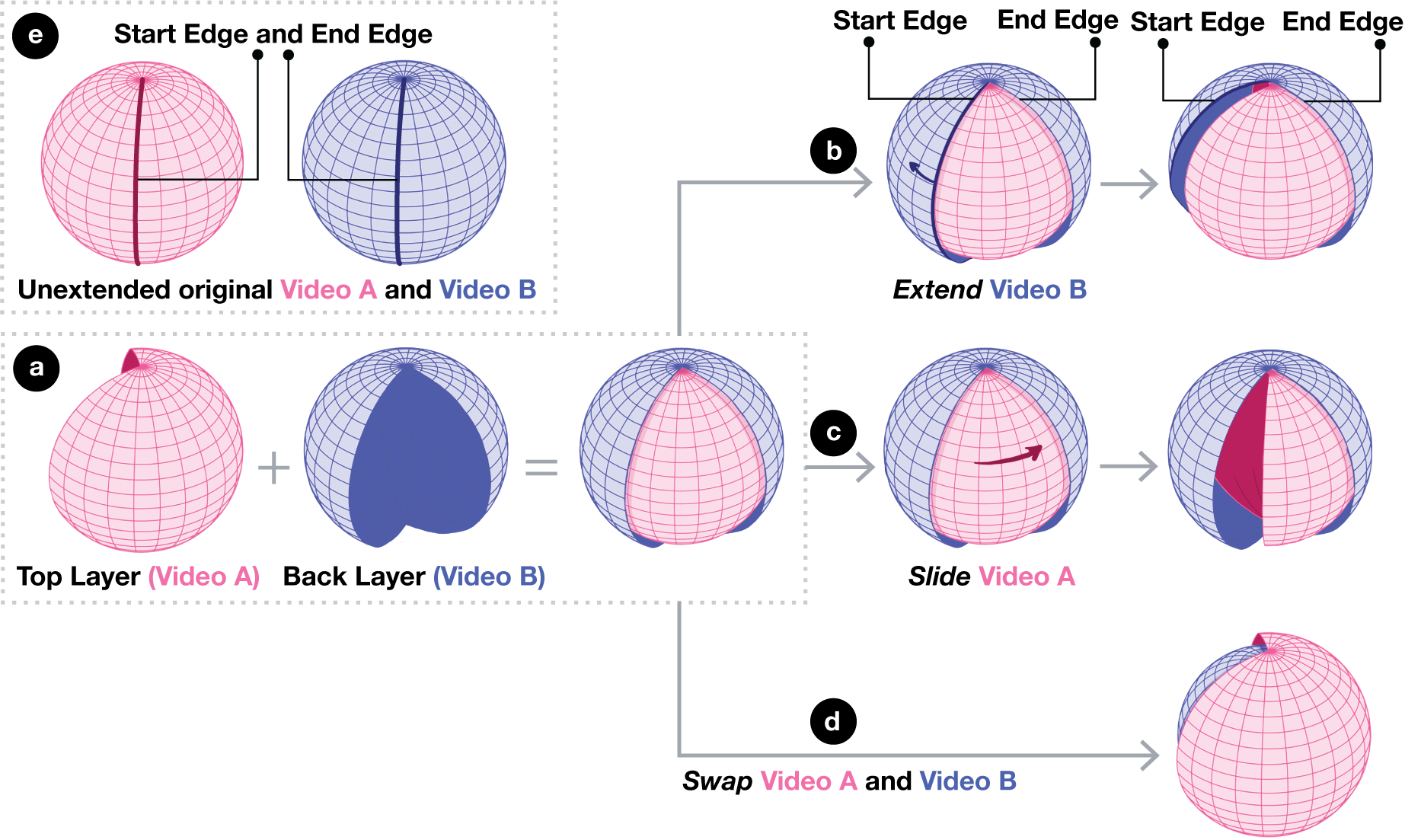}
    \vspace{-2mm}
    \caption{
Semantic illustration of our sliding concept in VR. 
In (e), the two original 360\textdegree{} IVs, Video~A (pink) and Video~B (blue), are shown as full spheres, with their start and end edges aligned. 
In (a), two spherical-lune IV display areas are combined. 
Panels (b)--(d) demo the three interactions supported by sliding: 
\textit{extend}\extend, which shrinks Video~B’s display area by moving its start edge (b); 
\textit{slide}\slide, which rotates Video~A so that the overlapping area gradually increases (c); 
and \textit{swap}\swap, which changes the front layer from Video~A to Video~B (d).
    }    
    \centering
    \label{fig:SlideInVR_semantic}
    \Description{
Semantic illustration of our sliding concept in VR. In (e), the two original 360-degree IVs, Video A (pink) and Video B (blue), are shown as full spheres, with their start and end edges aligned. In (a), two spherical-lune IV display areas are combined where Video A is the front layer, and Video B is the back layer. 
Panels (b)--(d) demo the three interactions supported by sliding: extend, which shrinks Video B’s display area by moving its start edge (b); slide, which rotates Video A so that the overlapping area gradually increases (c); and swap, which changes the front layer from Video A to Video B (d).
    }

    \vspace{6mm}

    \includegraphics[width=0.64\textwidth]{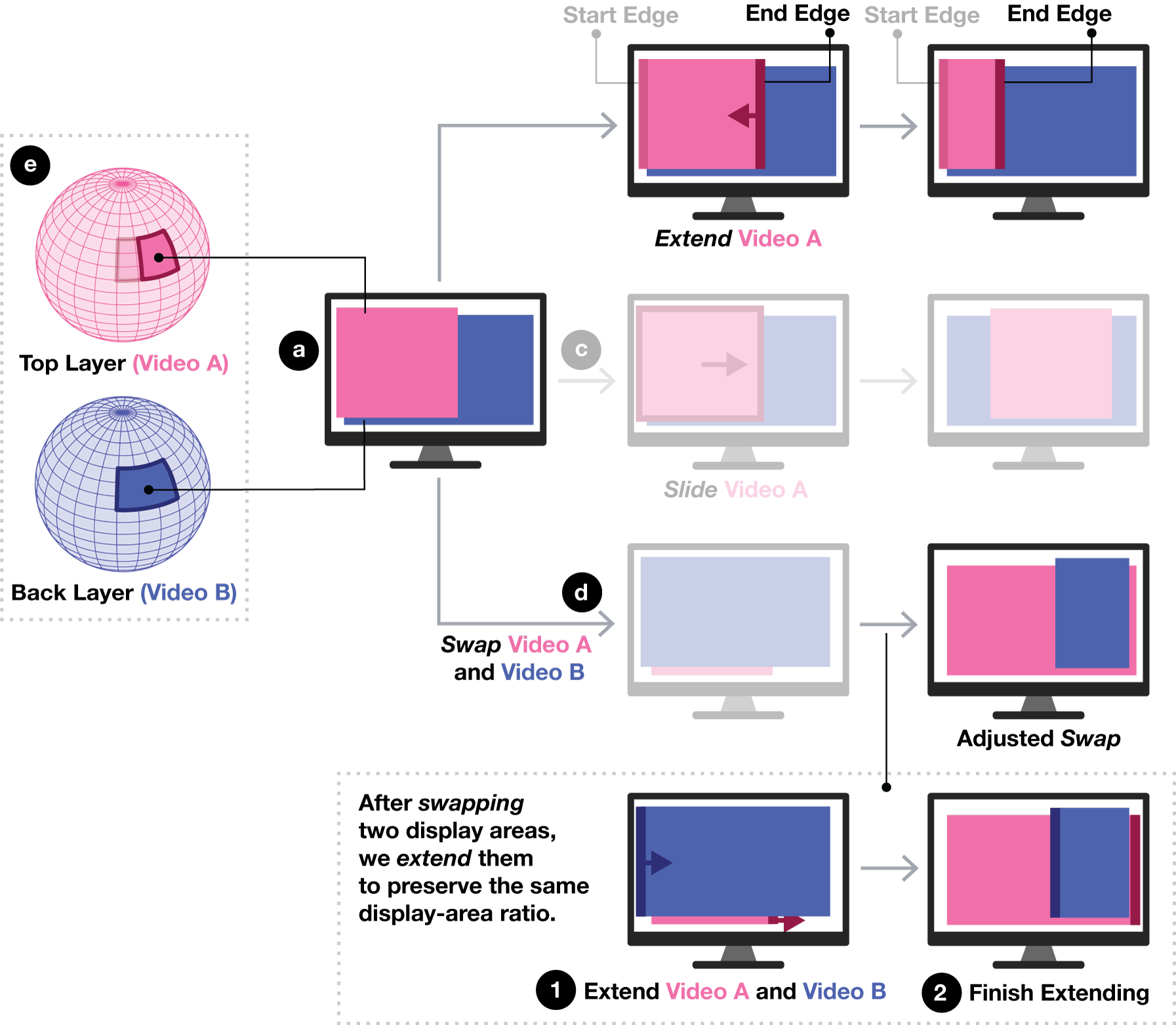}
    \vspace{-2mm}
    \caption{
Semantic illustration of our sliding concept in 2D. 
Partial original 360\textdegree{} IV in (e), Video~A (pink) and Video~B (blue), are shown in two overlapping IV display areas on the monitor in (a). 
Panels (b)--(d) demo the three interactions supported by sliding: \textit{slide}\slide (b), 
\textit{extend}\extend (c), and \textit{swap}\swap (d). To optimize monitor space and minimize interactions, during the implementations, we removed the \textit{slide}\slide (b) and made adjustments to \textit{extend}\extend (c) and \textit{swap}\swap (d), which will be explained in the \autoref{subsec:SlideIn2D}.
    }
    \centering
    \label{fig:SlideIn2D_semantic}
    \Description{
Semantic illustration of our sliding concept in 2D. 
Partial original 360-degree IV in (e), Video A (pink) and Video B (blue), are shown in two overlapping IV display areas on the monitor in (a). 
Panels (b)--(d) show the three interactions supported by sliding: slide (b), extend (c), and swap (d). To optimize monitor space and minimize interactions, during the implementations, we removed the slide and made adjustments to extend and swap, which will be explained in the section 4.3.
    }
\end{figure*}

\subsection{Adopting Sliding in VR}
Browsing one IV in VR is intuitive: the IV is projected onto the inside surface of an omnidirectional sphere, and the user is located in the center of the sphere to view the whole IV and navigate by rotating their VR headset.
When adopting sliding in VR, the two IVs are layered, with the top-layer IV ``nested inside'' the back-layer IV (\autoref{fig:SlideInVR_semantic}-a).
A start edge and an end edge of each IV---each resembling a great-circle arc on a sphere that passes through both poles---together define a spherical-lune display area, while the complementary region (from the end edge back to the start edge) is not rendered.
These edges determine which portion of each IV is displayed, \eg, $75\%$ (\autoref{fig:SlideInVR_semantic}-a) or $100\%$ when the two edges of an IV overlap (\autoref{fig:SlideInVR_semantic}-e). 
The size and placement of each IV display area can be adjusted through \textit{slide}\slide (\autoref{fig:SlideInVR_semantic}-b) and \textit{extend}\extend (\autoref{fig:SlideInVR_semantic}-c), and the layer order can be adjusted through \textit{swap}\swap (\autoref{fig:SlideInVR_semantic}-d).

\subsection{Adopting Sliding in 2D}
When viewing IVs in 2D environments, the IV is projected to a physical monitor.
As users cannot move the physical monitor, changing the FoV is commonly achieved by dragging the IV within a display area that shows only a partial view of the content.
Therefore, in 2D, each IV display area corresponds to the current FoV of that IV. 
When the sliding concept is adopted in 2D, two IV display areas are overlaid, with the front-layer IV rendered on top of the back-layer IV (\autoref{fig:SlideIn2D_semantic}-a).
Each IV display area has a left edge (start edge) and a right edge (end edge).
The size and placement of each IV display area can be adjusted through \textit{extend}\extend, which moves its edges (\autoref{fig:SlideIn2D_semantic}-b), and \textit{slide}\slide, which moves the entire display area (\autoref{fig:SlideIn2D_semantic}-c).
The layer order can be adjusted through \textit{swap}\swap (\autoref{fig:SlideIn2D_semantic}-d).
In our final implementation, we simplified this concept in 2D by removing the \textit{slide}\slide interaction and modifying \textit{extend}\extend and \textit{swap}\swap to better use screen space and reduce interaction steps, which will be explained in \autoref{subsec:SlideIn2D}.

\section{Technique Design and Implementation}

In this work, we aim to investigate how the two modalities---2D and VR---and the three comparison approaches---side-by-side, toggle, and sliding (\ie, flexibly supporting both side-by-side and toggle)---affect users' behaviors and experiences in comparing two IVs.
We design and implement five techniques: two applied the sliding concept (\ie, SlideInVR and SlideIn2D) and three other techniques (\ie, ToggleInVR, ToggleIn2D, and SideBySideIn2D).
Because full 360\textdegree{} IVs cannot be placed side by side in VR without distortion, we only implemented the side-by-side version in 2D (\ie, SideBySideIn2D).
We focus only on  IV comparison tasks in our study (\autoref{subsection:tasks}), rather than supporting a complete IV comparison workflow, which may include capturing, organizing, annotating, and editing IV contents.  
Thus, users do not define their own ROIs in our techniques. 
Instead, all techniques visualize one predefined ROI per IV, representing the primary area of interest in each IV.

For all techniques except SideBySideIn2D, the two IVs can be fully or partially overlaid; therefore, we provide a \textit{peek}\peek interaction~\cite{Baker:ComparingVideo:GI24}, which temporarily reveals a small region of the back-layer IV or the other IV.
All techniques offer a zoomed-in local view and a global minimap (azimuthal equidistant projection), visualize ROI trajectories, and include playback controls (play, pause, seek).

For techniques that support independent movement of each IV display area (\ie, SlideInVR, SlideIn2D, and SideBySideIn2D) , we implemented on-demand, ROI-based adjustments of the display areas to facilitate the comparison process.
To help users quickly distinguish the two IVs, we applied consistent color coding across techniques: pink for Video~A and blue for Video~B.

We used Unity (2022.3.16f1) and the Meta XR Interaction SDK (v66.0) to implement the two VR techniques (\ie, SlideInVR and ToggleInVR) and run the system on a Meta Quest~3, which has a 96\textdegree{} vertical field of view (vFoV) and a 104\textdegree{} horizontal field of view (hFoV). 
To implement the three 2D techniques, we used Vite and React together with Three.js for WebGL-based rendering of the different IV views.

\subsection{SlideInVR}
\begin{figure}[h]
    \includegraphics[width=0.47\textwidth]{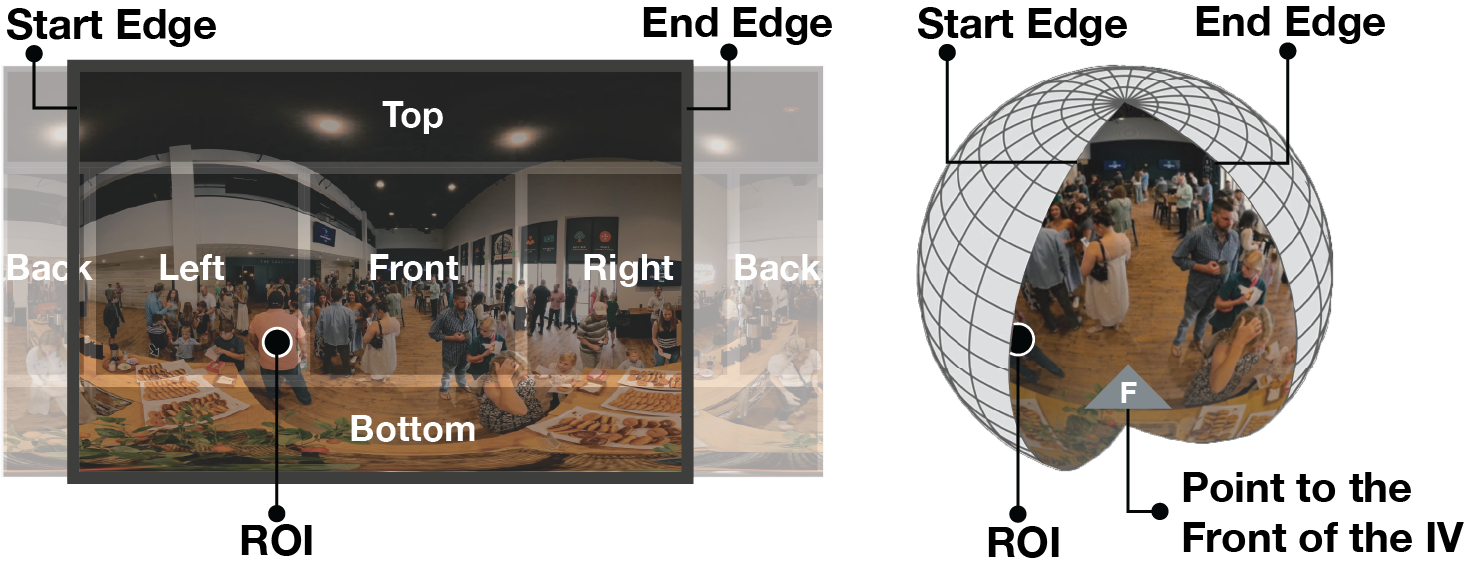}
    \vspace{-2mm}
    \caption{
    Mapping between the equirectangular IV frame and the spherical-lune display area in SlideInVR for the second tutorial task. 
    Left: a 360\textdegree{} IV \rev{~\cite{VR_church_video} from © Grace Church (used with permission) in equirectangular projection, annotated with directions (Front, Back, Left, Right, Top, Bottom) and the predefined ROI (the man in pink shirt).}
    The darker thick rectangle indicates the current display area.
    Right: the corresponding display area rendered on the lune sphere.
    }
    \centering
    \label{fig:SlideInVR_front_and_ROI_indicator}
    \Description{
    Mapping between the equirectangular IV frame and the spherical-lune display area in SlideInVR for the second tutorial task. Left: a 360-degree IV from © Grace Church (used with permission) in equirectangular projection, annotated with directions (Front, Back, Left, Right, Top, Bottom) and the predefined ROI (the man in pink shirt). The darker thick rectangle indicates the current display area. Right: the corresponding display area rendered on the inside of the lune sphere, with the arrow (F) indicating the front of the IV.
    }
    \vspace{-2mm}
\end{figure}

Following the sliding concept (\autoref{fig:SlideInVR_semantic}), SlideInVR enables users to view two IVs simultaneously in VR.
Custom shaders render both IVs in spherical-lune display areas (\autoref{fig:SlideInVR_front_and_ROI_indicator}) with corresponding minimaps, and users interact via VR controllers (Appendix \autoref{fig:control}-a).

\subsubsection{Extend, Slide, Swap, and Peek}
\begin{figure}[h]
    \includegraphics[width=0.47\textwidth]{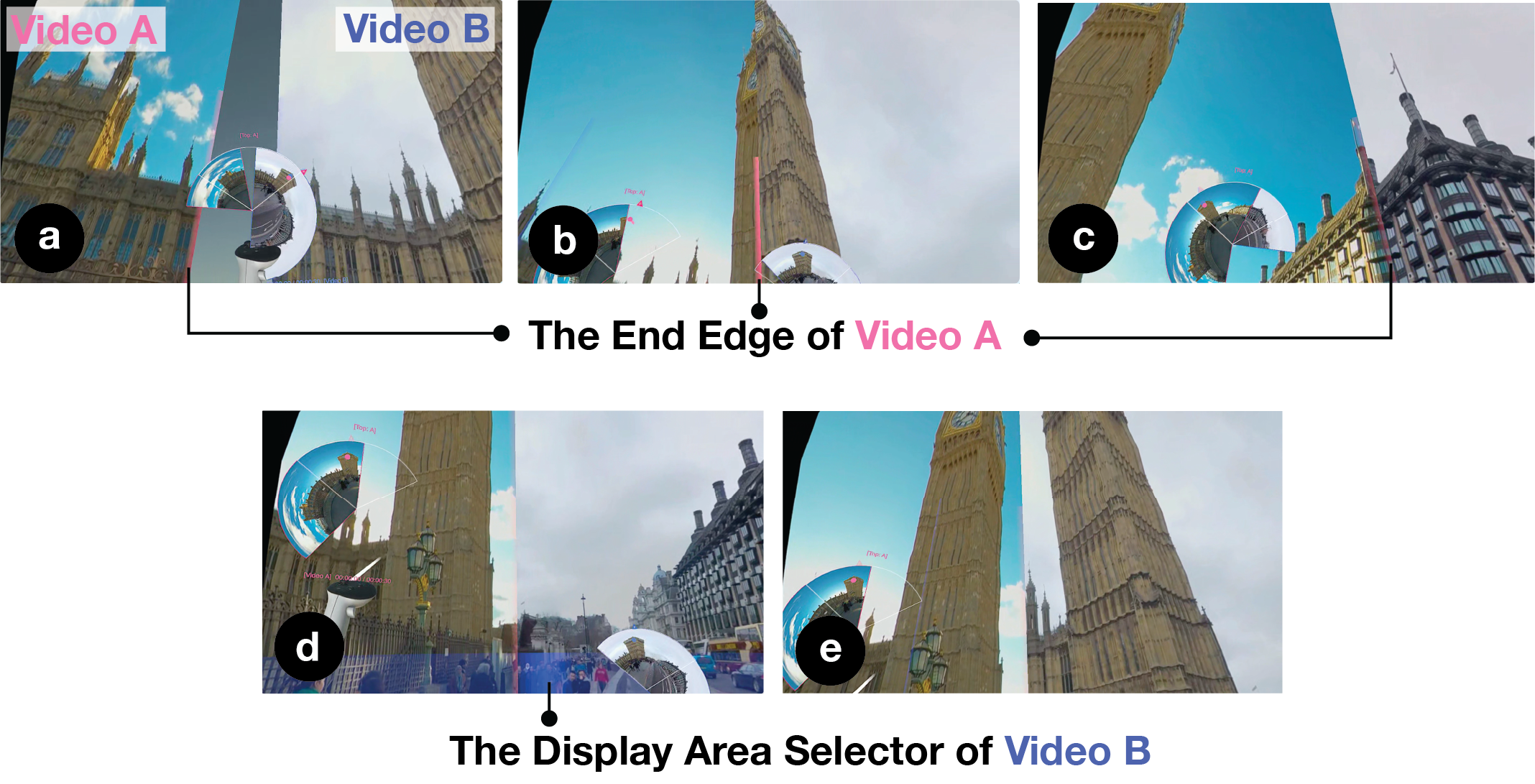}
    \vspace{-2mm}
    \caption{\rev{
    In (a), two display areas do not overlap, and the user plans to \textit{extend}\extend\ Video~A.
    In (b), the user is \textit{extending}\extend Video~A by moving the end edge to the right.
    In (c), the user stops \textit{extending}\extend, and the Big Ben in Video~A (top layer) is overlaid on the Big Ben in Video~B (back layer).
    In (d), the user is \textit{sliding}\slide the display area of Video~B by moving it clockwisely.
    In (e), the position of Video~B is updated once the user stops \textit{sliding}\slide.
    The two IVs are from © VR Gorilla, used with permission~\cite{VR_Gorilla_new_london,VR_Gorilla_old_london}.}
    }
    \centering
    \label{fig:SlideInVR_extend_slide}
    \Description{
    In (a), two display areas do not overlap, and the user plans to extend Video A. In (b), the user is extending Video~A by moving the end edge to the right. In (c), the user stops extending, and the Big Ben in Video A (top layer) is overlaid on the Big Ben in Video B (back layer). In (d), the user is sliding the display area of Video B by moving it clockwisely. In (e), the position of Video~B is updated once the user stops sliding.
    The two IVs are from © VR Gorilla, used with permission.
    }
    \vspace{-2mm}
\end{figure}

SlideInVR exposes four core interactions: \textit{extend}\extend, \textit{slide}\slide, \textit{swap}\swap, and \textit{peek}\peek.
With either controller, users grab and drag an edge to extend\extend an IV’s spherical-lune display area (\autoref{fig:SlideInVR_extend_slide}-b), and the size is updated in real time during extension.
The edges of the back-layer display areas are rendered thinner, helping users distinguish the order of layers and reducing visual distraction.
A band-shaped selector (pink above the equator for Video~A, blue below for Video~B) supports \textit{slide}\slide: 
users grab and drag the band to rotate the display area around the polar axis, with a gradient indicating the amount of sliding while updates occur on button release to reduce motion sickness (\autoref{fig:SlideInVR_extend_slide}-d,e).
Pressing any trigger button activates \textit{peek}\peek, temporarily revealing a small circular region (15\textdegree both vertically and horizontally) of the back-layer IV (\autoref{fig:SlideInVR_menu}-b).

\subsubsection{ROI-Based Display Manipulation}
\begin{figure}[h]
    \includegraphics[width=0.47\textwidth]{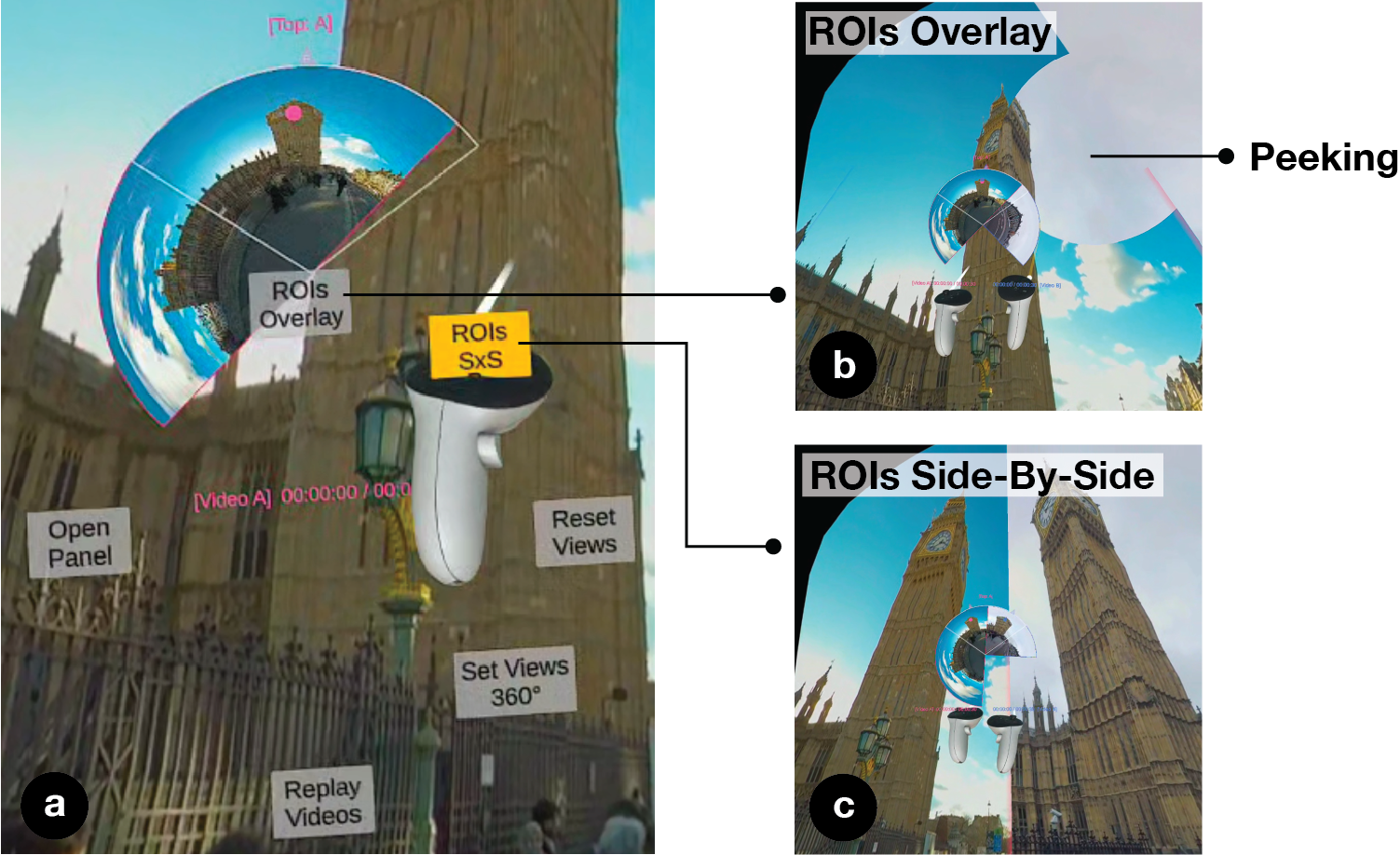}
    \vspace{-2mm}
    \caption{
    The hidden radial menu of SlideInVR advanced features surrounding the left controller in (a): \textit{ROIs Overlay} (top), \textit{ROIs SxS} (top right; selected and highlighted), \textit{Reset Views} (right), \textit{Set Views 360~\textdegree{}} (bottom right), \textit{Restart Videos} (bottom), and \textit{Open/Close Panel} (left). (b) ROIs SxS aligns the ROIs (Big Ben) side by side. (c) ROIs Overlay aligns the same ROIs overlaid. \rev{The two IVs are from © VR Gorilla, used with permission~\cite{VR_Gorilla_new_london,VR_Gorilla_old_london}.} 
    }
    \centering
    \label{fig:SlideInVR_menu}
    \Description{
    The hidden radial menu of SlideInVR advanced features surrounding the left controller in (a): ROIs Overlay (top), ROIs SxS (top right; selected and highlighted), Reset Views (right), Set Views 360-degree (bottom right), Restart Videos (bottom), and Open/Close Panel (left). (b) demonstrates that two ROIs are overlaid when the user triggers ROIs Overlay button, and (c) demonstrates that ROIs are placed side by side when the user triggers ROIs SxS button. The two IVs are from © VR Gorilla, used with permission.
    }
    \vspace{-2mm}
\end{figure}

A hidden radial menu provides ROI-based interactions and view presets (\autoref{fig:SlideInVR_menu}-a).
\btn{ROIs O} \btn{verlay} and \btn{ROIs SxS} automatically extend\extend and slide\slide the display areas so that the ROIs are overlaid (\autoref{fig:SlideInVR_menu}-b) or placed side by side (\autoref{fig:SlideInVR_menu}-c) in front of the user.
Additional items include \textit{Reset Views} (default half-and-half layout with Video~A on top), \btn{Set Views 360\textdegree{}} (full overlay layout, similar to ToggleInVR viewing mode), \textit{Restart Videos} (play all VIs from the beginning), and \textit{Open/Close Panel}.

\subsubsection{Minimap}
A minimap of Video~A or Video~B in azimuthal equidistant projection is attached above the left or right controller.
It shows the current display area, a FoV indicator, a front-direction arrow (``F''), and an ROI indicator (\autoref{fig:minimaps}-b).
SlideInVR also visualizes ROI trajectories on the minimap, with depth encoding time, and a 3D progress bar that provides better user awareness of video time~\cite{3D_timeline:Yu:2023} through the center to show video time.
The progress bar is hidden when the minimap faces the user to avoid occlusion and reappears when the controller is rotated.
When the IV is playing, the trajectory and progress bar move along the depth axis while the minimap itself stays fixed to the controller to reduce motion sickness.
Below the minimap, we show the current and total duration.
When the two controllers are close together, their minimaps, progress bars, and ROI trajectories overlay spatially, with the top-layer IV’s minimap rendered above the back layer.

\subsubsection{Temporal Navigation}
\begin{figure}[tb]
    \includegraphics[width=0.47\textwidth]{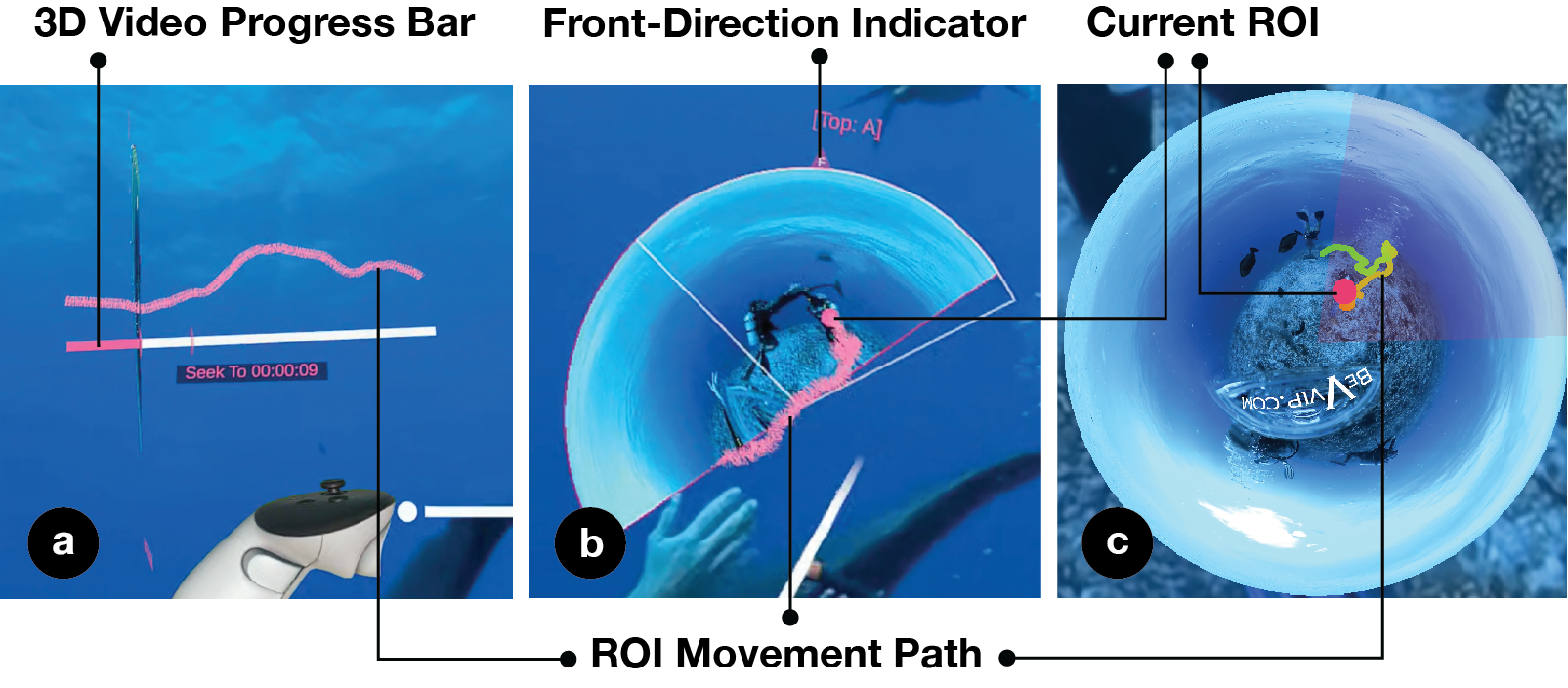}
    \vspace{-2mm}
    \caption {\rev{
    In (a), the user seeks within Video~A in SlideInVR by holding the ``X'' button and moving the left controller forward and backward.
    (b) demonstrates the spatial-temporal trajectory of the ROI of Video~A in SlideInVR;
    (c) demonstrates the gradient-color trajectory of the ROI of Video~A in any 2D techniques (from red to green).
    Example IV is from © Luxury Travel VR, licensed under CC BY 4.0~\cite{dolphin_video}. }
    }
    \centering
    \label{fig:minimaps}
    \Description{
        In (a), the user seeks within Video~A (a 3D video progress bar) in SlideInVR by holding the ``X'' button and moving the left controller forward and backward. (b) demonstrates the spatial-temporal trajectory of the ROI of Video~A in SlideInVR; (c) demonstrates the gradient-color trajectory of the ROI of Video A in any 2D techniques (from red to green). Example IV from © Luxury Travel VR, licensed under CC BY 4.0. 
    }
    \vspace{-2mm}
\end{figure}

Users navigate an IV using natural body movements. %
Navigation for Video~A and Video~B can be controlled independently using the left and right controllers, respectively, also allowing for navigating both IVs simultaneously.
They can \textit{seek} to a specific time in the corresponding  IV(\autoref{fig:minimaps}-a).
Moving the controller downward toggles play and pause for the corresponding IV.
In addition, moving the joystick forward or backward on either controller triggers a 5-second rewind or fast-forward, respectively, for the corresponding IV.

\subsection{ToggleInVR}
ToggleInVR displays a single IV on the entire sphere, similar to how users watch IVs in state-of-the-art systems.
With overlapping IVs, \textit{peek}\peek reveals a small region of the other IV, and ``B'' toggles between them.
The minimap, progress bar, and spatially visualized ROI trajectory follow the same design as in SlideInVR.
However, only the current IV’s minimap is visible, and the projection always shows its full overview.
All other interactions for activating advanced features and navigation are the same as in SlideInVR, except that users can only navigate the current viewing video.
In the hidden menu, only \textit{Restart Videos} (bottom) and \textit{Open/Close Panel} (left) are provided.
No ROI-based advanced features are supported because ToggleInVR presents only one IV at a time, and users are not able to reposition the IV.
A full overview of the interaction mappings is provided in Appendix \autoref{fig:control}-b.

\subsection{SlideIn2D}
\label{subsec:SlideIn2D}

All 2D-based techniques follow the same interface: 
a row of clickable features, the canvas for displaying IVs, and then temporal navigation controls as shown in Figures~\ref{fig:SlideIn2D} and \ref{fig:SideBySide_Toggle_In2D}.

\subsubsection{Physical Navigation}
In SlideIn2D, the canvas consists of two display areas.
Each IV is shown in its own display area, with the highlighted top-layer display area (partial) overlaid on the back-layer display area.
Each IV viewport has a fixed vFoV of 40\textdegree, while the hFoV is adjusted according to the size of the browser and the size of the display area.
In our implementation, the maximum hFoV of SlideIn2D is $\approx98$\textdegree, derived from the symmetric perspective frustum of THREE.js:
\[
\begin{gathered}
hFoV = 2 \cdot \arctan\!\left(\tan\!\left(\tfrac{vFoV}{2}\right)\cdot\tfrac{W}{H}\right),\\
\text{where $W/H$ is the aspect ratio of the display area.}
\end{gathered}
\]

\begin{figure*}[tb]
    \includegraphics[width=0.7\textwidth]{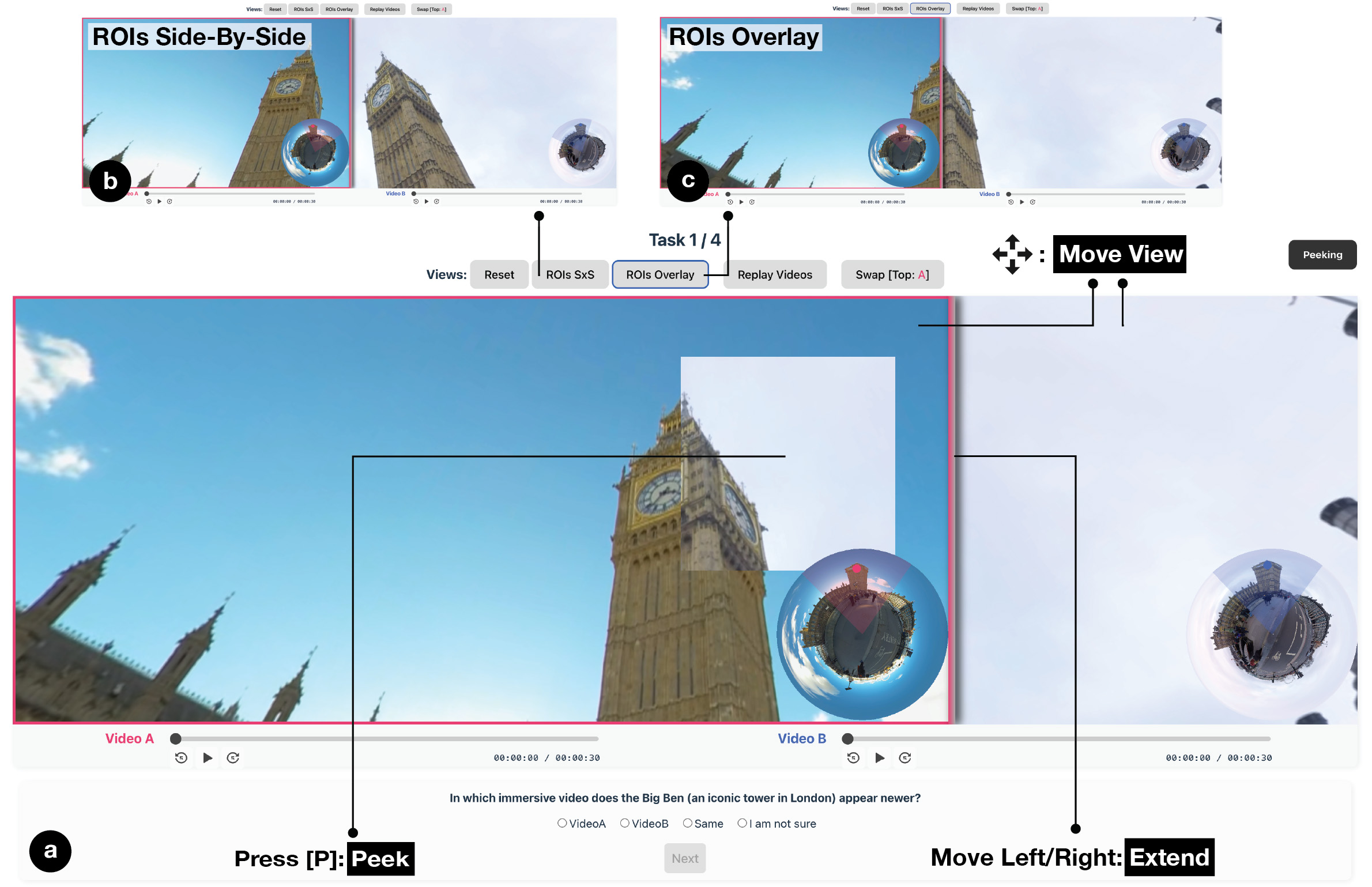}
    \vspace{-2mm}
    \caption{
    \rev{(a) Overview of SlideIn2D showing two IVs~\cite{VR_Gorilla_old_london,VR_Gorilla_new_london} from © VR Gorilla (used with permission) and core controls. (b) ROIs SxS aligns the ROI (Big Ben) side by side. (c) ROIs Overlay aligns the same ROI for overlay-based comparison.}
    }
    \centering
    \label{fig:SlideIn2D}
    \Description{(a) Overview of SlideIn2D showing two IVs from © VR Gorilla (used with permission) and core controls. (b) ROIs SxS aligns the ROI (Big Ben) side by side. (c) ROIs Overlay aligns the same ROI for overlay-based comparison.}
\end{figure*}

Users can drag the display area to navigate the corresponding IV (\autoref{fig:SlideIn2D}), where horizontal dragging adjusts the yaw of the viewpoint and vertical dragging adjusts the pitch.
They can then use the \textit{Reset} button to return both IVs to their default front-facing views.

\subsubsection{Extend, Swap, and Peek}
Users drag the slider (one edge of the top-layer IV) left or right to \textit{extend}\extend its display area and reallocate horizontal space (\autoref{fig:SlideIn2D}-a).
To simplify the screen real estate and simplify interactions, SlideIn2D instantiates only a subset of sliding features: users can extend only the top-layer IV via a single edge, while the other edges are fixed to keep the back-layer IV fully visible (\autoref{fig:SlideIn2D_semantic}), ensuring back-layer display area is always fully visible.
Thus, the back layer cannot be \textit{extended}\extend or \textit{slid}\slide.
Users can \textit{swap}\swap the layer order to foreground the other IV.
Pressing ``P'' activates \textit{peek}\peek, temporarily revealing a $300\text{px} \times 300\text{px}$ (about 21\textdegree{} $\times$ 21\textdegree{}) region of the back-layer IV at the mouse position; a “Peeking” label in the upper-right corner indicates this mode.

\subsubsection{ROI-Based Display Manipulation}
We provide two features for automatically adjusting both IV views based on the current ROI positions.
\btn{ROIs SxS} recenters each ROI within its display area, placing them side by side on the canvas (\autoref{fig:SlideIn2D}-b), while \btn{ROIs Overlay} recenters both views so the ROIs overlap on the canvas (\autoref{fig:SlideIn2D}-c).

\subsubsection{Minimap}
In the corner of each display area, a minimap shows an azimuthal equidistant overview of the IV, with the current ROI position and a gradient-colored trajectory (red to green) depicting its motion over time, following prior designs~\cite{Time_Curves_trajectory} (\autoref{fig:minimaps}-c).

\subsubsection{Temporal Navigation}
Below the IV display areas, temporal controls for Video~A (left) and Video~B (right) allow users to drag a progress bar to seek, as in a traditional video player.
Additional controls include play/pause, 5-second rewind/forward, current/total duration, and a \textit{Replay Videos} button to restart both IVs.

\begin{figure*}[tb]
    \includegraphics[width=1\textwidth]{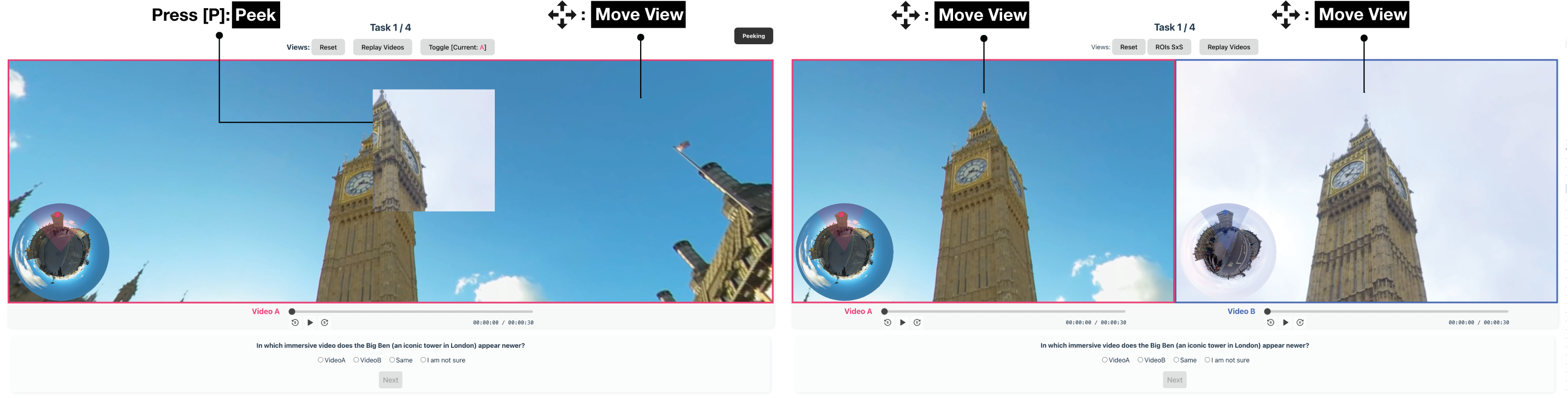}
    \vspace{-6mm}
    \caption{The overview of ToggleIn2D (Left) and SideBySideIn2D (Right) \rev{showing two IVs from © VR Gorilla (used with permission)~\cite{VR_Gorilla_old_london,VR_Gorilla_new_london}.}}
    \centering
    \label{fig:SideBySide_Toggle_In2D}
    \Description{The overview of ToggleIn2D (Left) and SideBySideIn2D (Right) showing two IVs of Task\#14. (© VR Gorilla. Used with permission.)}
    \vspace{-2mm}
\end{figure*}

\subsection{ToggleIn2D}
ToggleIn2D (\autoref{fig:SideBySide_Toggle_In2D}) presents a single FoV that shows one IV at a time. 
The view (40~vFoV and 98\textdegree{} hFoV) is shared by both IVs, so dragging the view updates the viewpoint of both videos simultaneously.
Users can reset to the default front-facing view with \textit{Reset} and switch between IVs using the \textit{Toggle} button (\autoref{fig:SideBySide_Toggle_In2D}).
Because a single shared view applied to both IVs, ROI-based advance features, including \btn{ROIs SxS} and \btn{ROIs Overlay}, are not supported.
Since the IVs are fully overlaid, \textit{peek}\peek temporarily reveals a small region of the other IV.
Navigation controls and minimaps follow SlideIn2D, but users can navigate only the currently visible IV, consistent with ToggleInVR.
\textit{Replay Videos} restarts both IVs.

\subsection{SideBySideIn2D}
SideBySideIn2D (\autoref{fig:SideBySide_Toggle_In2D}) splits the canvas into two equal views, each with 40\textdegree{} vFoV and 59\textdegree{} hFoV: left for Video~A and right for Video~B.
Users can drag each view independently to navigate its IV.
\btn{ROIs SxS} automatically centers each ROI within its FoV.
Because views never overlay, \textit{peek}\peek and \btn{ROIs Overlay} are not supported. 
The minimap and temporal navigation controls follow the same design as in SlideIn2D.

\section{User Study}

\subsection{Participants and Apparatus}
We recruited 20 participants (eight females and 12 males) where eight were aged 18–24, and 12 were aged 25–34.
Recruitment was conducted via university mailing lists, posts in relevant local VR Discord communities, and word of mouth. 
We screened for VR and immersive video (IV) viewing experience and excluded individuals with no prior experience.
Participants reported diverse VR experience (from $<1$ year to $>5$ years) and IV viewing frequency (Appendix~\autoref{fig:freq}).
Most participants (14 out of 20) had at least one year of VR experience, while 11 out of 20 participants reported using VR occasionally ($N=7$), regularly ($N=2$), or frequently ($N=2$).
In contrast, the majority (16 out of 20) reported watching IVs rarely ($N=5$) or very rarely ($N=9$).
Participants received 20~CAD per hour as a remuneration, and the study was approved by our institutional research ethic office.

All participants completed study tasks on a wheeled office chair.
For the three 2D techniques, they used a 28-inch monitor with a keyboard and mouse, and for the two VR techniques, they wore a Quest~3 HMD and used two controllers.

\subsection{Tasks}
\label{subsection:tasks}

We designed four domain-independent IV comparison tasks [T1--T4] that were derived based on previous work on low-level comparison tasks in information visualization~\cite{Amar:Low-Level_Components_of_Analytic_Activity_in_Information_Visualization:2025} and extended from 2D video comparison tasks~\cite{Baker:ComparingVideo:GI24} (\autoref{tab:task_types}):
\begin{itemize}
  \item[\textbf{[T1]}] \textbf{Identify differences in the temporal occurrences of frames and ROIs.} 
For example, a coach may compare two 360\textdegree{} training recordings of the same drill to determine in which attempt a player enters a key position earlier, performs a move more often, or stays in a critical area for longer.
  
  \item[\textbf{[T2]}] \textbf{Compare spatial-temporal motion of ROIs.}  
  For example, an IV film director or editor may watch two trials of a scene and compares the actors’ movements and transition of postures~\cite{Murch:Blink:2021,JumpCut_CinematicVR_30and180-degree:2024}.
  
  \item[\textbf{[T3]}] \textbf{Find visual differences in the visual attributes of ROIs, such as size, color and static posture.} 
  For example, a historian or environmental scientists may view and compare IVs captured at different times to identify the changes of building appearances or surrounding environments~\cite{historical_tour:Abidin:2020, climate_change_comparison:Markowitz:2018}. 

  \item[\textbf{[T4]}] \textbf{Identify the visual difference between the full immersive video frames.} 
  For instance, an AI scientist may compare video graphics or panorama quality of two AI-generated panorama videos to determine whether frames in one IV are more distorted or more consistent and have less stitches at the edges~\cite{360DVD:2024}.
\end{itemize}

\begin{table*}[ht]
\centering
\small
\caption{Four types of comparison tasks with abstract question templates and selected actual questions.}
\label{tab:task_types}
\resizebox{1\textwidth}{!}{%
\begin{tabular}{c p{9cm} p{6cm}}
\toprule
Task Type & Abstract Question Templates & Selected Actual Questions \\
\midrule

{[T1]}& 

  \begin{itemize}
    \item In which IV, does the [ROI] appearing or disappearing happen first?
    \item In which IV, does the [ROI] in red color appear more or less frequently?
    \item Does one IV have a accumulated longer duration for the [ROI] (\eg, an object staying in view longer before exiting)?
  \end{itemize} & 
  \begin{itemize}
  \item In which immersive video does the person in white greet earlier?
  \item In which immersive video does the shark attack more frequently?
\end{itemize}  \\
\hline

{[T2]} & 
  \begin{itemize}
    \item In which IV, does the [ROI] move toward or away from a reference or an area (\eg, the front area of an IV)?
    \item In which IV, does the [ROI] moves faster or slower?
    \item In which IV, does the [ROI] (\eg, a person) change posture or facial expression more frequently or more dynamically?
  \end{itemize} &
  \begin{itemize}
      \item In which immersive video does the diver wave the paddle more consistently?
      \item In which immersive video does the lemur (indicated by the arrow) move along a more circular path?
  \end{itemize} \\
\hline

{[T3]}& 
  \begin{itemize}
    \item Are the two [ROIs] the same object or person?
    \item In which IV, does the [ROI] show more diversity in appearance (\eg, color or shape)? 
    \item In which IV, does the [ROI] (\eg, a person) exhibit a more theatrical posture or facial expression at a specific moment?
  \end{itemize} & 
    \begin{itemize}
      \item Are the two female lions near the camera in each immersive video the same individual?
      \item In which immersive video does the Big Ben (an iconic tower in London) appear newer?
  \end{itemize} 
 \\
\hline
{[T4]} & 
      \begin{itemize}
        \item Which IV has better graphics quality (\eg, less noticeable distortion, blurring and stitches)?
        \item Which IV has better frame consistency?
        \item Do the two IVs depict the same environment or location?
      \end{itemize} & 
      \begin{itemize}
      \item Do the two immersive videos depict the same location?
      \item Which immersive video has better quality (e.g., less noticeable distortion, stitches or blurring)?
  \end{itemize} \\
\bottomrule
\end{tabular}
}
\end{table*}

\subsection{Study Data: Immersive Videos Clips}
We selected 20 pairs of 30-second 360\textdegree{} IV clips for the within-subject comparison tasks, covering four comparison tasks [T1--T4] across five techniques.
Each pair was associated with a meaningful question requiring participants to compare and identify video content (Appendix~\autoref{tab:appendix_tasklist}).

Most study stimuli were natural, real-world IVs sourced from YouTube: 18 of the 20 pairs were natural clips.
We additionally included 2 pairs of simulated AI-generated IVs specifically for T4, because T4 focuses on assessing global visual differences (\eg, distortions, blurring, and stitching artifacts) that are difficult to obtain with controlled severity from in-the-wild videos.
Due to limitations of current generative AI (GenAI) models for IVs (\eg, graphics quality)~\cite{360DVD:2024}, we simulated ``AI-generated'' IVs by re-recording real scenes in Unity and applying controlled filters to introduce different levels of distortion, blur, and stitching artifacts.

Across 20 pairs, clips were drawn from (i) the same environment and perspective, (ii) the same environment but different perspectives, or (iii) entirely different environments.
We also selected ROIs exhibiting diverse visibility patterns: ROIs may remain visible throughout an entire video, disappear and reappear in the same or different locations (\eg, in clips with transitions), or appear and disappear multiple times.
To reduce ambiguity in identifying the intended ROI, six clips were augmented with a brief arrow indicator in the first five seconds to highlight visually similar ROIs.

For each [T1--T3] task , we pre-defined exactly one ROI per IV based on the corresponding question.
For ROI visualization, we used a pre-trained open-source 360\textdegree{} tracking model\footnote{https://github.com/VitaAmbroz/360Tracking
} to track and annotate the chosen ROI in each IV clip.
In clips where the ROI disappeared and reappeared (\eg, with transitions), separate tracking segments were combined into a single trajectory. 
No ROI tracking and visualizations were performed for [T4], which required assessing global visual differences rather than comparing localized content.

\subsection{Procedure}
Participants provided informed consent and completed a demographic pre-survey.
We then introduced 360\textdegree{} IVs, azimuthal equidistant projection, and four comparison tasks [T1--T4].
Next, participants completed all five comparison techniques (\ie, SlideInVR, ToggleInVR, SlideIn2D, ToggleIn2D and SideBySideIn2D) in a balanced Latin square order. 

In each condition, participants received a brief slide-based introduction to the technique interactions, followed by a short tutorial with four practice tasks using the same four IV pairs, and were encouraged to explore their own comparison strategies.
After the tutorial, participants performed four experimental tasks in random order, each corresponding to a unique, non-repeated IV pair from the 20 prepared pairs (Appendix~\autoref{tab:appendix_tasklist}).
After completing four tasks, participants filled out a post-questionnaire on their perceptions of the technique and strategies used.
Participants could take short breaks between conditions to reduce fatigue.

After all five conditions, participants took part in a 15-minute semi-structured interview to discuss their perceptions of five techniques, preferences, workflows, and strategies for comparing IVs.
The study lasted an average of 2.27 hours per participant (min = 1.70 and max = 2.77).
We conducted the study in a single session to keep participants’ task and technique experiences fresh for supporting more reliable comparisons.

\subsection{Measures}
During each comparison task, we recorded participants' performance, including accuracy.
In the post-questionnaire, participants reported their strategies and rated their perceptions of task workload and comparison techniques on 7-point Likert scales.
These ratings included standardized instruments (NASA-TLX~\cite{nasa}, UMUX-Lite~\cite{umux-lite}) as well as custom self-defined questions.
In the interview, participants ranked the five techniques by dragging their names on a slide and provided rationales for their rankings.

\subsubsection{Qualitative Analysis}
In addition to quantitative measures, we also conducted a thematic analysis of participants’ free-text responses in the post-questionnaires and the transcripts of the semi-structured interviews. 
The first author, who ran all the study, generated an initial codebook related to (i) workflows of IV comparison, (ii) comparison strategies (\eg, toggle vs. side-by-side, parallel vs. sequential), and (iii) preferences and perceived benefits and drawbacks of each technique and modality.
The first two authors then independently coded the responses and transcripts by assigning relevant excerpts to these codes and adding, splitting, or modifying codes when new patterns emerged (\ie, a hybrid inductive–deductive approach). Finally, the same two authors grouped related codes into higher-level themes, resolving disagreements through discussion; a third author acted as a tiebreaker when needed.

\section{Results}
In this section, we report how modality and comparative approach shaped participants’ workflows, strategies, and subjective experience when comparing IVs.
Our analysis combines qualitative observations and interviews with quantitative ratings and accuracy data.
Across techniques, we did not observe significant differences in task accuracy (Section~\ref{sec:quantitative_results}); instead, the main effects of the techniques were reflected in how participants navigated, compared, and experienced the tasks.

\subsection{Overall Workflows}
Across techniques, participants followed a two-phase workflow during IV comparison: navigation and comparison.
Participants first moved within the 360\textdegree{} space to ``find,'' ``locate,'' ``navigate,'' or ``track'' predefined or self-defined ROIs and then applied the suitable comparison strategies.
For [T1--3], most participants typically located the ROI(s) in one or both IVs and then compared using parallel or sequential viewing (Section~\ref{sec:comparison}).
For [T4], participants often treated the task as iterating [T3] multiple times.
Some participants also created their own anchor points to support global comparison (\eg, P18 \qt{tried to manually select anchors and then compare for a fair comparison for global visual comparison.}).

\subsubsection{Phase 1: Spatial Navigation}
In the study, participants showed different workflows with different techniques.
In SlideInVR, SlideIn2D and SideBySideIn2D, participants heavily used \btn{ROI SxS} and \btn{ROIs}\hspace{0.2em}\btn{Overlay} features to automatically find the ROIs before comparing.
Many described \btn{ROI SxS} as a favorite feature because it reduced search time and effort to locate ROIs (P1,7-9,11,12,14,16,20).
\pqt{I liked the ROIs side by side (ROIs SxS) a lot, that was the MVP. It just fixes everything for you, as long as the ROI isn't moving that much.}{P11}

After initial ROI alignment, we observed two common behaviors: some participants continued using ROI-based view adjustment to follow moving ROIs, while others preferred manual dragging to maintain spatial context in 2D.
\pqt{When I was dragging (in 2D), I got a little more understanding about the scenery around me ... If I directly only rely on the button. I feel like it kind of locks me into one of viewpoint, and I have no understanding.}{P20}

For ToggleIn2D and ToggleInVR, because the two IVs share the same view, there was no way to make the two ROIs overlaid or side by side.
Thus, participants heavily relied on the minimap and manual navigation to locate ROIs.
P8 described, \qt{they don't have ROIs overlay, which means that you need to have a lot of navigation.}

\subsubsection{Phase 2: Comparison}
Once ROIs were located, the technique strongly shaped comparison style.
Participants described SlideIn2D and SlideInVR as the most ``powerful'' and ``flexible'' because they supported fluid switching between side-by-side and overlay views, along with space reallocation via extending\extend.
As P2 summarized, \pqt{SlideInVR is the one with more possibility.}{P2}

By contrast, SideBySideIn2D enforced persistent side-by-side viewing, while ToggleIn2D and ToggleInVR often pushed participants toward sequential viewing, especially when ROIs moved or appeared in different locations.
As P9 described,
\pqt{Toggling both in VR and in 2D was somewhat horrible for tracking motion. When objects are moving in a very different way or show in drastically different locations, toggling back and forth doesn't let you track both. You still have to go through each video individually.}{P9}

The peek\peek feature partially mitigated this limitation by providing a concurrent local view when ROIs were not aligned, enabling a form of real-time side-by-side comparison (P3,6):
\pqt{For ToggleInVR, I find the peeking was extremely helpful. Without it I had to toggle all the time and could not compare in real time. But with the peeking, I could just compare the point of interest simultaneously.}{P3}

\subsection{Physical and Temporal Navigation} \label{sec:navigation}

We next describe how techniques shaped navigation, both physically (viewpoint changes in 360\textdegree{} space) and temporally (timeline control).

\subsubsection{Physical Navigation}
Few participants found 2D navigation easier due to less body movement and familiar mouse input (P5,7,8), whereas the majority described VR navigation as more natural and intuitive (P1-7,10-12,14,16-18,20) especially for shifting attention across directions or following fast-moving ROIs (P1,5,7,13,16,20). 
As P18 described, \qt{for the shark (fast-moving ROIs), in 2D, it was very hard to keep track of the ROI, but in VR it would have been much easier because I can just turn my head a bit.}

VR also fostered a stronger sense of spatial understanding through embodied interaction (P9,10,14–16,19). 
As P14 described, \qt{... on a flat screen (2D techniques), I can forget where things are because I drag to get there. But in VR, I just know it's back there, and I just look there.}
In addition, the larger FoV in VR reduced the need for manual navigation during comparison, because participants could capture more relevant content with ``a little bit'' (P8) head rotations.

However, the larger FoV in VR also increased visual information and could raise cognitive demand (P3,8,9).
In addition, some participants reported a ``dizziness'' feeling (P9) and disorientation during full-scene toggling in ToggleInVR (P9,11,15). 
\rev{As P15 noted,
\qt{... when it completely changes, I have that brief half second moment was like, `where am I?' And then I have to focus on it again... At least the SlideInVR lets you control how much changes, because there's that idea of multiple sections.}}

Many participants found the minimap helpful for locating and verifying ROIs, especially for moving targets (P4-7,11,14,15,17,20) and when using 2D techniques (P11,14).
As P20 reflected,
\qt{The lemur jumped off the screen, and I couldn’t figure out whether it was the same one sitting there or another one. Then I had to look at the minimap and see that I was following the wrong lemur.}
Some also used the minimap for comparing distances or paths through visualized trajectories (P4-6,12,15,18), setting up display areas in SlideInVR (P3) and providing a global overview of either the entire frame (P6,19) or large ROIs (P10,18,19).

\subsubsection{Temporal Navigation}
Several participants reported that temporal control felt more intuitive in VR, allowing play/pause/seek with less attentional shift (P1,5,8,12). 
For instance, P1 noted, 
\qt{For the 2D, I had to pause or play when I'm focusing on what I'm doing on the screen, It takes more effort and it causes more delay... But in VR... I can just unintentionally use my hands to do that motion, and I don’t have to put my attention towards my hands and the replay buttons.}

\subsection{Comparison Strategies} \label{sec:comparison}
Participants’ strategy choices were strongly task-dependent rather than uniformly tied to a particular technique. 
Almost all participants (except P10) noted that they used or would use different comparison approaches depending on the task.
In this section, we describe how different task characteristics shaped participants’ preferences and the strategies they employed.

\subsubsection{Side-by-side vs. Toggle}
In most tasks, many participants preferred to use side-by-side (P1,4-7,9,11,12,15,16), as perceived as \qt{a natural and straightforward comparison method that they are already familiar with} (P3,6,14-16).
They also explained that side-by-side reduced the fear of \qt{missing something} (P7-9,11,14,16) and lowered memory workload by eliminating the need to recall details (P8,11) since both videos remained visible during comparison.
As P11 noted, \qt{it is beneficial, because sometimes if you're doing it one at a time, you're relying on your brain to remember everything correctly, and sometimes it just doesn't. But seeing them side by side sometimes reveals information that you might have missed.}
Side-by-side also supported time efficiency (P3,6,14) and was effective for comparing time steps or tracking differences across temporal sequences (P3) by playing two IVs simultaneously.

Toggle was often preferred for fine-grained visual differences (\eg, subtle shape or quality differences) (P3,6,8,9,12,15-18).
Reflecting on a [T3] task, P8 stated: 
\qt{I remember the lion posture task clearly. I used the overlay (toggle) very quickly, because it gave me very clear differences between how [the lion posture in] the top layer is different from [the lion posture in] the second layer.}{P8}
This approach was especially preferred when comparing ROIs with similar shapes appearing in the same location or perspective (P1,5-8,13,16,19,20).
However, toggle became ineffective when ROIs were moving or when videos were captured from different perspectives, and participants switched to side-by-side (P1-3,5-7,9,16,18-20).
As P16 summarized, \qt{for example, for the video quality in the same location, peek (toggle) works quite nicely. But for a lot of the videos that actually don't match together, peeking (toggling) a specific region wouldn't show much difference.}
These results align with Yuan~\etal’s findings that for detecting subtle local and global visual differences in static 360\textdegree{} images, toggle is preferred and perceived as more effective than placing ROIs side by side.

Overall, no single approach was sufficient across all tasks.
Participants therefore stressed the importance of supporting flexibility in switching between comparison approaches (P1-4,6,8,11,12,15-19). 
As P11 noted, \qt{each of these tasks would shine in one of the tools ... some of them are just better if you watch them side by side ... some of them are very good if you can use a peek with the overlay (toggle).}

\subsubsection{Parallel vs. Sequential}
Participants exhibited two multitasking approaches: parallel and sequential viewing.
While some favored parallel (P9,11,14,15) and others preferred sequential (P7,13,19), most reported that their choice depended on the task.
Participants used parallel viewing by rapidly shifting attention when differences were subtle (P3,12), ROIs were static (P7,16,17), or videos are long (P12,14).
As P12 explained, \qt{the good thing is the video is short, so you don’t need to memorize a lot ... But when there are many things to remember with only tiny differences, like in video A I took six donuts, but in video B I took seven, it’s really hard to memorize. That’s why I don’t prefer watching only one video at a time.}

However, when ROIs were moving substantially or quickly, participants switched to watching IVs sequentially (P7,9,11,13,16,17,20).
As P17 reflected on a [T2] task:
\qt{If the movement is very mild, then parallel is so good... but the lemur was very hard because it was moving a lot, and I had to drag my mouse to move my view to follow it. 
}
Besides tasks that demanded more physical movements, participants also preferred sequential multitasking when tasks imposed high mental workload, such as memorization, counting, or comparing multiple objects (P3,7,9,12-14,16,18,20).
As P3 explained: \qt{For really, really simple tasks I can still view the two videos at the same time. But if it’s a bit harder, like if I need to count the donuts or compare how many times the shark attacks, I need to carefully watch one video after another. In that case, I would prefer to use toggle, or even if they are side by side, I would only play one video at a time.}

\subsection{Quantitative Results}
\label{sec:quantitative_results}
We report quantitative results for perceived helpfulness, workload, usability, and task accuracy. 
Because Likert and NASA-TLX ratings deviated from normality (Shapiro--Wilk, $p < .05$), we used Friedman tests with Wilcoxon signed-rank post-hoc tests with Holm--Bonferroni correction.
Trial-level accuracy (binary) was analyzed using a mixed-effects logistic regression with participant as a random effect.
In line with common HCI practice, we treat $p < .05$ as evidence against the null hypothesis and consider larger $p$-values as descriptive rather than inferential.

\begin{figure*}[tb]
    \includegraphics[width=1\textwidth]{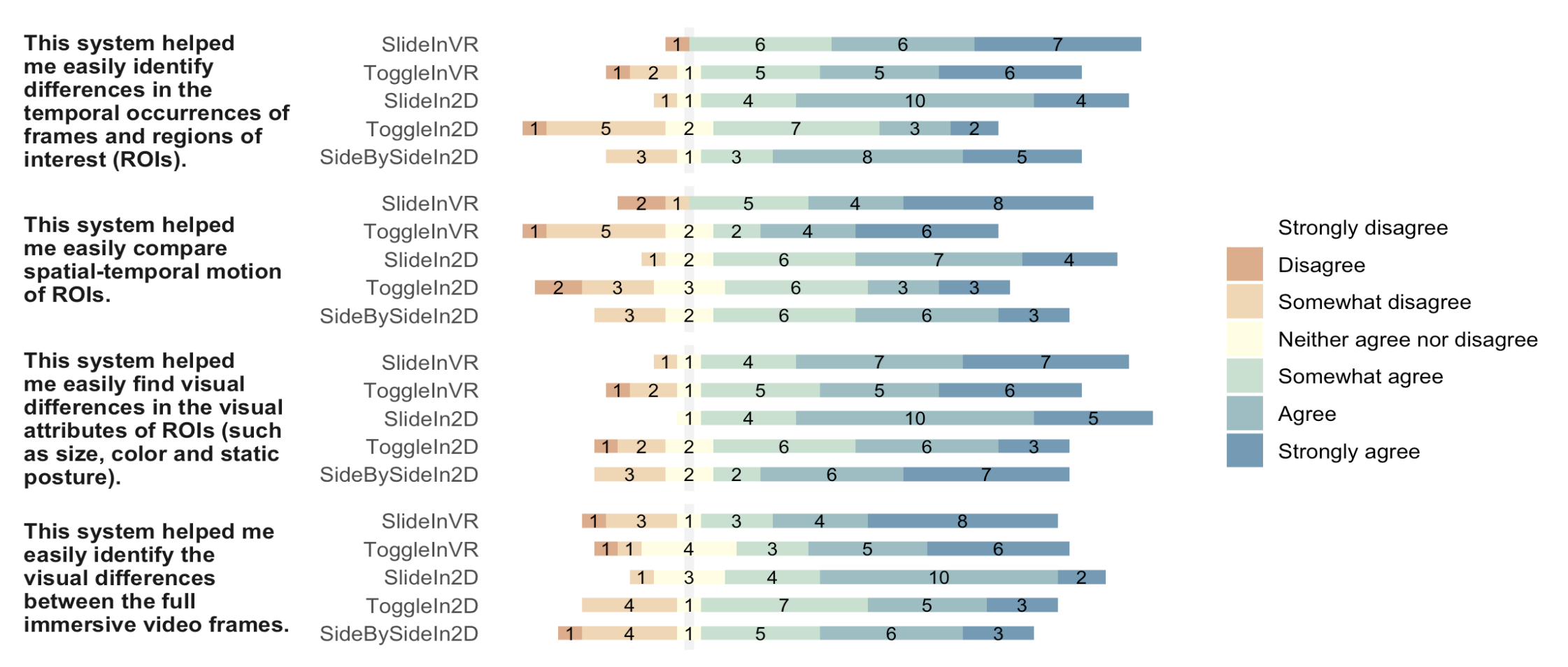}
    \vspace{-8mm}
    \caption{Stacked bar charts of how participants thought that each comparison techniques was helpful for four types of comparison tasks.}
    \centering
    \label{fig:helpfulness_t1-4}
    \Description{Stacked bar charts of how participants thought that each comparison techniques was helpful for four types of comparison tasks.}
    \vspace{-2mm}
\end{figure*}

\subsubsection{Helpfulness}
Participants rated how helpful they found each technique for each task type~\autoref{fig:helpfulness_t1-4}. 
For [T1] (temporal occurrence), a Friedman test indicated an omnibus effect of Technique ($p = .017$), but no pairwise comparisons reached significance ($p = .064$) after Holm--Bonferroni correction.
For [T2] (spatial location), [T3] (visual difference), and [T4] (event sequence), Friedman tests did not reveal significant differences in helpfulness across techniques ($p \ge .19$).
Detailed statistical results and additional rating distributions are provided in Appendix~\autoref{tab:helpfulness_stats} and~\autoref{fig:7-likert-others}.

\begin{figure*}[t]
    \includegraphics[width=1\textwidth]{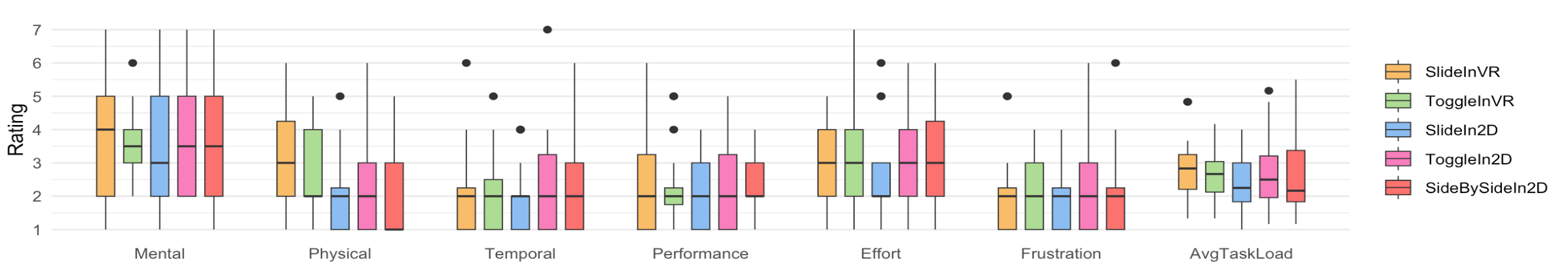}
    \vspace{-6mm}
    \caption{Box-and-whisker diagrams of unweighted NASA-TLX ratings on a 7-point Likert scale (the lower the better).}
    \centering
    \label{fig:nasa_combine}
    \Description{Box-and-whisker diagrams of unweighted NASA-TLX ratings on a 7-point Likert scale (the lower the better) for five techniques.}
\end{figure*}

\subsubsection{Workload}
We collected participants' perceptions of their experience using the NASA-TLX questionnaire (\autoref{fig:nasa_combine}).
Friedman tests reveal no significant differences for most subscales or average ratings, except for physical demand ($p < 0.001^{***}$).
Post-hoc tests did not yield statistically significant pairwise differences for physical demand, including SlideInVR vs. SlideIn2D ($p = .072$), SlideInVR vs. ToggleIn2D ($p = .061$), and SlideInVR vs. SideBySideIn2D ($p = .072$). 
Descriptively, mean physical demand ratings for SlideInVR were higher than for the 2D techniques (Appendix~\autoref{tab:tlx_subscales}), which is consistent with participants’ reports that VR required more physical effort (\eg, head and body movements). 

These results are also reflected in the interviews.
Participants appreciated its flexibility and power, but many also described it as demanding to use. 
Many participants reported that SlideInVR imposed a high workload to operate, remember and recall its controls (P2,4,8,10,12,15,20). 
As P2 explained: 
\qt{SlideInVR is the one with more possibility, but it’s the most overwhelming... SlideInVR is amazing, but the workload is too high. Maybe if I was working with it for a couple of weeks, that would be automatic.}
Participants also reported difficulty in distinguishing and managing the two video layers (P5,7), which requires more mental and physical effort during IV comparison.
As P5 noted,
\qt{For SlideInVR, it’s really not that easy to tell apart which layer is on top [layer]. When you’re looking at a video, you don’t know which layer it is without moving it away to see what is below.}

\subsubsection{Usability}
We assessed perceived usability with the UMUX-Lite scale.
The Friedman test did not reveal a significant difference in usability ratings across the five techniques ($p = .08$).
Descriptively, SideBySideIn2D tended to receive the highest median usability ratings and ToggleIn2D the lowest (Appendix~\autoref{tab:umux_combined}), but these differences did not meet our criterion for statistical significance.

\subsubsection{Accuracy}
In our accuracy analysis, ``I am not sure'' responses (2.75\% of answers) were treated as incorrect, as they indicate that participants could not determine which IV better satisfied the comparison criterion.
Item-level inspection (\autoref{tab:appendix_tasklist}) showed that three tasks (8, 10, and 19) had the lowest accuracy (30–40\%). 
Per-technique accuracy on these tasks also varied considerably, and some Technique~$\times$~Task combinations were represented by only 2–7 trials, reflecting the limited sample size and randomized task assignment.

We modeled trial-level accuracy using a mixed-effects logistic regression with Technique and TaskType as within-subject fixed effects and a random intercept for participant. 
The model revealed no significant main effect of Technique ($p \ge .313$), no significant main effect of TaskType ($p \ge .19$), and no significant Technique $\times$ TaskType interaction ($p \ge .24$). 
Mean accuracies ranged from $0.64$ to $0.75$ across techniques (Appendix~\autoref{tab:accuracy}), suggesting that, within our study, the different techniques and task types did not lead to systematic differences in correctness.

\subsection{Learnability and Preferences}

\subsubsection{Learning Curve}
In general, participants perceived 2D techniques and controls as more familiar, while VR required additional adaptation.
As P9 described:
\qt{I think if I grew up using VR, then I’d feel less comfortable with 2D. But for now, it’s a little bit hard for me to adapt [to VR].}
This challenge was particularly evident in SlideInVR. 
Although we provided sufficient time for participants to familiarize themselves with each technique, most participants reported that SlideInVR involved a steep learning curve (P1–3,5,7,8,10,12,13,15,17,18).
Many described feeling overwhelmed at first, as they had to grasp a new interaction concept while also learning to operate numerous features.
As P12 explained:
\qt{I think that (SlideInVR) is a promising technique, but for me, it's a little bit overload, because I didn't have past experience working with that, and I spent a really long time to figure out which button I should press, or how can I achieve my goal. But I think that is a decent technique to help me to catch the differences between two videos.}

\subsubsection{User Preference}
\begin{figure}[tb]
    \includegraphics[width=0.47\textwidth]{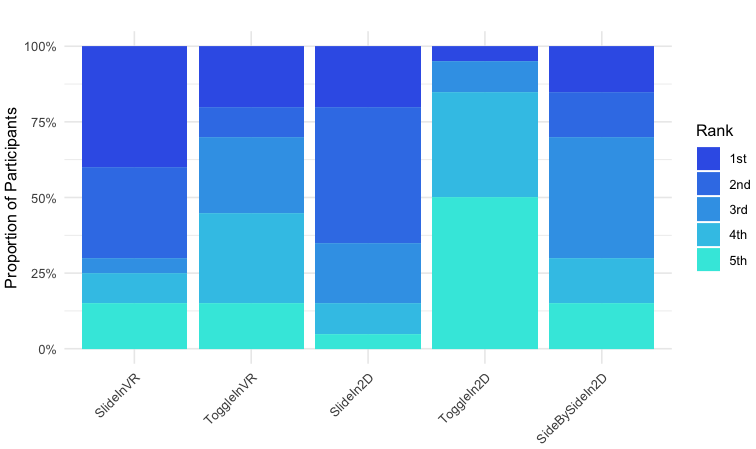}
    \vspace{-4mm}
    \caption{Stacked bar charts of comparison technique rankings from the most favorite (1st) to the least favorite (5th).}
    \centering
    \label{fig:ranking}
    \Description{Stacked bar charts of comparison technique rankings from the most favorite (1st) to the least favorite (5th).}
    \vspace{-2mm}
\end{figure}

For the overall experience, most participants ranked SlideInVR and SlideIn2D as their top two favorite comparison techniques, and ToggleIn2D as the least.
Specifically, SlideInVR received $40\%$ first-place votes and $70\%$ top-2 placements; SlideIn2D received $65\%$ top-2 placements, while ToggleIn2D was ranked last by $50\%$ of participants, as shown in Figure~\ref{fig:ranking}.
A Friedman test reveals a significant effect of technique on preference ($p < 0.001^{***}$). 
The post-hoc pairwise tests using Wilcoxon signed rank test with Holm–Bonferroni correction indicate a significant difference in preference between SlideInVR and ToggleIn2D ($p = 0.04 ^{*}$), SlideIn2D and ToggleIn2D ($p = 0.006 ^{**}$), and SideBySideIn2D and ToggleIn2D ($p = 0.04 ^{*}$).

In addition, participants’ explanations clustered around several recurring rationales: 
powerfulness (P1,5,6,8,9,11,14-20) and flexibility (P1,3-6,9,10,12-15,17,19,20);
ease of use (P3-5,7,8,11-20);
familiarity with the technique (P2,4,7,10-13) and its input devices (P7,12,20);
learning curve (P2,3,8,12,13,18);
effectiveness of comparison (P3,8,12,19,20);
naturalness of navigation (P1,4,6,11-14,16-18);
workload (P1,2,4,7,8,16,17,20); and
immersive (P5) and intuitive (P9,10,15,16) IV viewing experience brought by VR.

Many participants perceived VR as better supporting comparisons of graphics quality and distortion than 2D (P1,3,6,8-10,12,16,19, 20).
This advantage relates to the WYSIWYG nature of VR, where distortions and artifacts felt more immediate and noticeable.
As P6 explained:
\qt{It is more natural if you see a distortion in the VR, and you feel that fake immediately.}
Participants also emphasized that VR’s immersive qualities and sense of presence made subtle differences easier to detect.
As P19 described:
\qt{In VR, everything I can view more carefully. I think I’m a part of it so I’m able to view things just better... Texture is more clear in the VR, I’m able to understand things better just because there’s more depth to it.}
\rev{Similarly, beyond VR-specific benefits, some participants also valued SlideIn2D for allowing adjustable screen-space allocation during side-by-side comparison (P3,7,9,11,15,16,20).
As P15 pointed out,
\qt{SideBySideIn2D is already an upgrade... but sometimes I don’t really care about a video that much. So I want to make sure that I’m looking at 80\% of one video, instead of 50/50 all the time. So it was nice to have that option, and the good thing about SlideIn2D is that even though there was an extra option, it felt very natural and it didn’t add mental demand.}}

In summary, our results indicate that IV comparison follows a two-phase workflow: users first navigate to relevant ROIs and then apply comparison strategies.
During navigation, users' perceptions were shaped by different modalities and techniques:
VR made spatial and temporal navigation feel more natural and helped participants judge quality and distortion, but it also came with higher physical demand and a steeper learning curve.
Participants’ preferences for comparison strategies were strongly task-dependent.
They switched between side-by-side and toggle viewing, and between parallel and sequential viewing, based on how fine-grained the comparison was, where and how the ROI moved, task complexity, and memory demands, rather than relying on a single approach.
SlideInVR and SlideIn2D were perceived as more powerful and flexible because they enabled fluid switching between side-by-side and toggle-style comparisons, whereas ToggleIn2D was consistently the least preferred.
Within the scope of our study, these differences in workflows, strategies, and user experience did not translate into systematic accuracy differences across techniques and task types, suggesting that the primary benefits of the techniques lie in how they shape comparison processes rather than raw performance outcomes.

\section{Discussion}

\subsection{When Sliding Helps (and When It Does Not)}
Our study indicates that supporting both side-by-side and overlay-style comparison is important for IV comparison tasks.
The sliding concept provides a flexible way to move between these comparison approaches, enabling users to adjust how much of each video is visible and to choose the comparison style that best fits the current task.
\rev{Note that our study was conducted only on tasks comparing two IVs, and thus future studies might be needed to assess the insights below for multiple IV comparison.}

In 2D, sliding helped participants make better use of limited screen space and attention by allocating more area to the view that was most important for the current comparison.
Participants also found sliding particularly useful when tracking moving ROIs, since they could adjust how much of each video was visible instead of relying on a fixed 50/50 split.
Although we used a relatively large monitor (28 inches), several participants still used sliding to hide less relevant content and focus on key regions or fast movements.
We believe these benefits to be even more pronounced on smaller displays.

In VR, the sliding concept helps to reduce the physical movements of heads and bodies to switch between different views during comparison tasks, especially for those objects located in different locations. 
\rev{Also, as P15 and several other participants reported, it could help reduce large visual changes and motion sickness when comparing IVs in VR.}

These qualitative findings help explain why sliding techniques were often preferred and perceived as powerful and flexible, even though our quantitative analysis did not show clear accuracy differences between techniques. 
However, SlideInVR is still a novel prototype that introduces a multi-layer representation of IVs.
On the one hand, having multiple display areas can partially break immersion, as participants’ attention is drawn to the interface rather than the scene itself.
On the other hand, our current design only supports comparing two IVs at a time, and many participants already found it difficult to distinguish the layer order in VR. 
Extending sliding to more than two simultaneous IVs in VR would therefore require additional design work to avoid visual clutter and confusion, and the benefits are not guaranteed~\cite{360_image_comparison:Yuan:2025}.

In addition, sliding primarily supports \emph{how} users allocate visual attention, manage motion, and control transitions, rather than directly improving correctness on short, well-defined comparison questions.
For designers, this suggests that sliding is most valuable when users must repeatedly reframe views, track moving ROIs, or work within limited visual space, whereas simpler techniques may remain sufficient for short, low-workload comparisons.

\subsection{Technique-Specific Trade-offs}
\rev{In order to understand how different approaches (\ie, toggle, side-by-side, and combined) support IV comparison, we implemented five different techniques in our study. In the following, we synthesize our key findings into technique-level strengths and limitations, which could shed light on future technique development.}

\begin{itemize}
    \item \textbf{SlideInVR and SlideIn2D} offered the greatest \emph{flexibility} and perceived \emph{power}, allowing participants to fluidly switch between side-by-side and toggle comparison and to reallocate space for ROIs.
    Their ROI-based advanced features also facilitated faster navigation.
    This flexibility allowed them to support a wide range of comparison tasks and made them particularly suitable for more complex or unfamiliar IV comparisons that required switching between different comparison strategies. 
    They were strongly preferred overall, but the flexibility came with higher workload and a steeper learning curve, especially in VR.
    Their advantages therefore manifested primarily in richer workflows and user preference, rather than in significantly higher accuracy.
    \item \textbf{SideBySideIn2D} provided a simple, familiar, and usable baseline for many tasks.
    It supported continuous side-by-side comparison and was perceived as efficient and low-risk in terms of ``missing something,'' but lacked the ability to easily switch to overlay-style comparisons.
    Thus, SideBySideIn2D was particularly suitable for easy, low-workload, or long-duration comparison tasks, as well as for comparing trends in relative large monitor.
    \item \textbf{ToggleIn2D and ToggleInVR} were effective for detecting fine-grained local differences when ROIs were aligned in space, but became cumbersome when ROIs moved or scenes differed substantially.
    ToggleInVR also introduced disorientation due to abrupt scene changes.
    Participants rarely preferred these techniques overall, despite their usefulness for specific tasks.
    Thus, both ToggleIn techniques were particularly suitable for comparing short IVs captured from the same camera or for high-workload, detail-focused comparison tasks.
\end{itemize}

\section{Limitations and Future Work}
Our study examined how users adopted strategies across techniques and tasks, revealing both the promise and current challenges of five IV comparison techniques. 
We next highlight key limitations and opportunities for future research.

First, although we designed tasks to be as objective as possible, some comparisons remained inherently subjective and were interpreted differently across participants. 
For example, in Task\#13, participants disagreed on whether a lion using its paw to hold the meat constituted a distinct eating posture. 
In Task\#10, participants also varied in how they interpreted ``more circular'' (\eg, a more geometrically circular shape vs.\ a more complete circle). 
\rev{Such interpretive differences made some questions more difficult, increased task performance time and limited strictly objective performance evaluation.}

Second, participants frequently relied on ROI side-by-side and ROI overlay features to quickly locate relevant content, suggesting strong potential for more autonomous ROI support. A promising direction is to integrate AI that can pre-analyze videos (\eg, track multiple ROIs that may intermittently appear and disappear) and adapt layouts based on task context and users' prior interaction patterns. This could reduce manual tool switching, better maintain ROIs in view, and lower workload during immersive comparison.

Third, our work focused on monoscopic 360\textdegree{} IVs and primarily on \emph{pairwise} comparison between two clips at a time. Future work should explore how these techniques generalize to other IV formats (\eg, stereoscopic 360\textdegree{} and 180\textdegree{}), as well as how interaction and layout should scale to \emph{multi-video} comparison scenarios involving three or more IVs.

Finally, our evaluation used short (30-second) real-world IV clips with a single predefined ROI per clip. Future studies could validate the techniques in domain-specific workflows (\eg, training, filmmaking, editing, game scene design, and AI-generated IV analysis), and examine richer ROI pipelines where users can annotate, refine, and manage multiple ROIs over longer and more complex videos.

\section{Conclusion}
In this work, we adapt the concept of \textit{sliding} from 2D image and video comparison and apply it to VR and IV comparison, enabling users to flexibly switch between the side-by-side and toggle comparative approaches.
We \rev{implement} five comparison techniques across 2D and VR modalities and investigated them in a user study involving four types of comparison tasks. 
We also examined how different techniques and modalities shaped users’ workflows and strategies. 
Our findings show that users preferred and applied different techniques depending on the task, with SlideInVR and SlideIn2D rated as the two most favored techniques. 
These two techniques were valued for their powerfulness and flexibility, as they supported smooth switching between comparative approaches and provided better control for allocate space. 
At the same time, our results reveal challenges, particularly with SlideInVR, which highlights opportunities for refining interaction design to reduce complexity, manage workload, and improve comfort.

\begin{acks}
This work is supported in part by the Natural Sciences and Engineering Research Council of Canada (NSERC) Discovery Grant \#RGPIN-2020-03966, the Canada Foundation for Innovation (CFI) John R. Evans Leaders Fund (JELF) \#42371, and gift funds from Meta and Cisco.
We acknowledge that much of our work takes place on the traditional territory of the Neutral, Anishinaabeg, and Haudenosaunee peoples. Our main campus is situated on the Haldimand Tract, the land granted to the Six Nations that includes six miles on each side of the Grand River.
\end{acks}

\bibliographystyle{ACM-Reference-Format}
\bibliography{references/references.bib, references/hci_methods.bib, references/video_list}
\clearpage

\appendix
\onecolumn
\section{Appendix}
\begin{figure*}[ht]
    \includegraphics[width=0.9\textwidth]{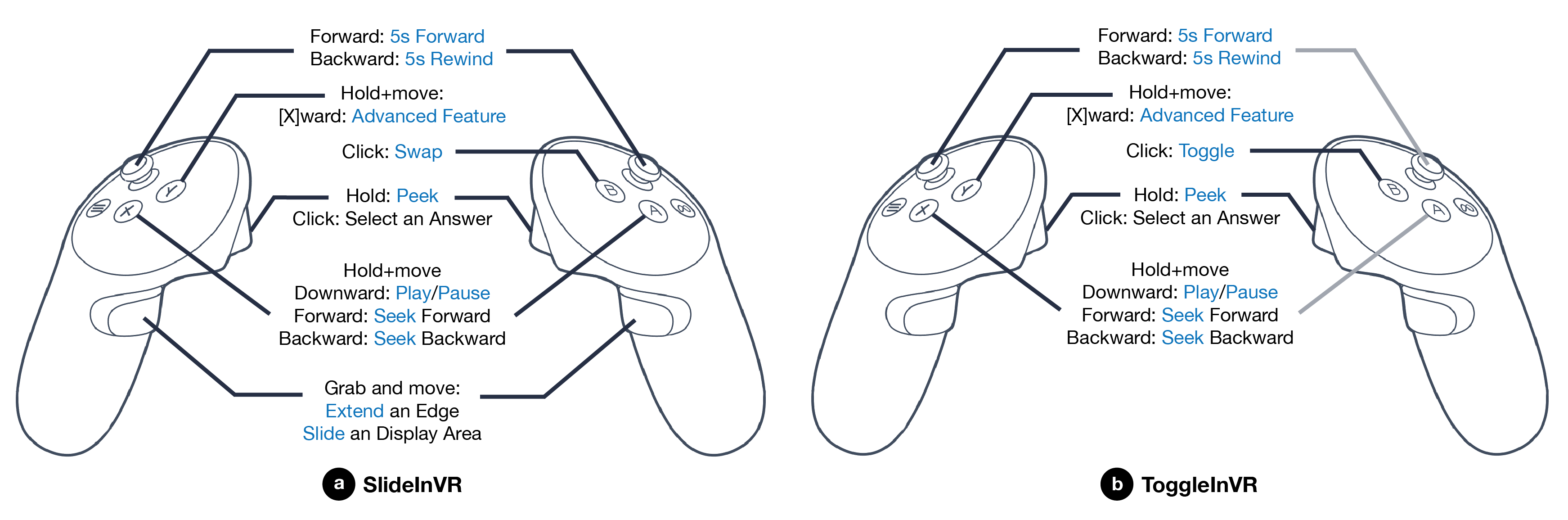}
    \vspace{-2mm}
    \caption{An overview of VR controller interactions in SlideInVR (a) and ToggleInVR (b), illustrating button functions.}
    \centering
    \label{fig:control}
    \Description{An overview of VR controller interactions in SlideInVR (a) and ToggleInVR (b), illustrating button functions.}
    \vspace{-2mm}
\end{figure*}

\begin{figure*}[ht]
    \includegraphics[width=1\textwidth]{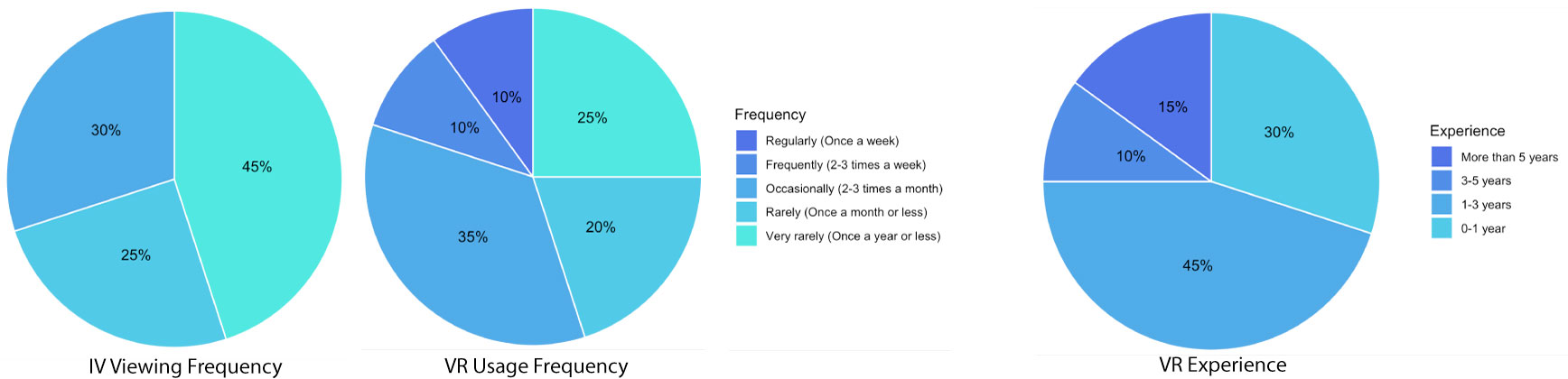}
    \vspace{-4mm}
    \caption{Three bar charts showing distribution of participants' IV viewing frequency (left), VR usage frequency (middle), and VR experience (right).}
    \centering
    \label{fig:freq}
    \Description{Three bar charts showing distribution of participants' IV viewing frequency (left), VR usage frequency (middle), and VR expereince (right).}
\end{figure*}

\begin{table*}[ht]
\centering
\caption{A full task list used in the user study.
Each task contains task ID, task type, comparison question, options of the answer, correct answer, whether two videos were applied with filters, and average accuracy in the study.
Each video lasts for 30 seconds.}
\label{tab:appendix_tasklist}
\resizebox{\textwidth}{!}{%
\begin{tabular}{c c p{7cm} p{4cm} c c c}
\toprule
TaskID & TaskType & Question & Options & Answer & Applied Filters & Avg. Accuracy \\
\midrule
1  & T1 & In which immersive video does the person in white greet earlier? & [VideoA, VideoB, Same, I am not sure] & VideoB &  & 100\% \\
\hline

2  & T1 & In which immersive video does the shark attack more frequently? & [VideoA, VideoB, Same, I am not sure] & VideoA & & 85\% \\
\hline

3  & T1 & In which immersive video does the lemur (indicated by the arrow) spend more time facing away from the viewer? & [VideoA, VideoB, Same, I am not sure] & VideoA & & 50\% \\
\hline

4  & T1 & In which immersive video does the bear (indicated by the arrow) stop more frequently? & [VideoA, VideoB, Same, I am not sure] & VideoA &  & 55\%\\
\hline

5  & T1 & In which immersive video do more white cars than black cars pass by Sagrada Familia (a famous church in Spain)? & [VideoA, VideoB, Same, I am not sure] & VideoB &  & 75\%\\
\hline

6  & T2 & In which immersive video does the female lion move closer to the observer camera for a longer duration? & [VideoA, VideoB, Same, I am not sure] & VideoA &  & 80\%\\
\hline

7  & T2 & In which immersive video does the aurora move faster in the sky? & [VideoA, VideoB, Same, I am not sure] & VideoB &  & 80\%\\
\hline

8  & T2 & In which immersive video does the cloud move faster in general in the sky? & [VideoA, VideoB, Same, I am not sure] & Same &  & 30\%\\
\hline

9  & T2 & In which immersive video does the diver wave the paddle more consistently? & [VideoA, VideoB, Same, I am not sure] & VideoA &  & 70\%\\
\hline

10 & T2 & In which immersive video does the lemur (indicated by the arrow) move along a more circular path? & [VideoA, VideoB, Same, I am not sure] & VideoA &  & 40\%\\
\hline

11 & T3 & In which video do the fireworks show significanly more diverse or vibrant color changes? & [VideoA, VideoB, Same, I am not sure] & VideoB &  & 70\%\\
\hline

12 & T3 & Are the two female lions near the camera in each immersive video the same individual? & [Yes, No, I am not sure]  & No &  & 45\%\\
\hline

13 & T3 & Does the male lion eat the meat in a similar posture in two immersive videos? & [Yes, No, I am not sure]  & Yes &  & 80\%\\
\hline

14 & T3 & In which immersive video does the Big Ben (an iconic tower in London) appear newer? & [VideoA, VideoB, Same, I am not sure] & VideoA &  & 65\%\\
\hline

15 & T3 & In which immersive video does the aurora appear more intense at its peak (e.g., brighter)? & [VideoA, VideoB, Same, I am not sure] & VideoB &  & 85\%\\
\hline

16 & T4 & Do all the people who appear in immersive video A also appear in immersive video B, even if their accessories (e.g., headwear) may differ? & [Yes, No, I am not sure]  & Yes &  & 80\% \\
\hline

17 & T4 & Do the two immersive videos depict the same location? & [Yes, No, I am not sure]  & Yes &  & 95\% \\
\hline

18 & T4 & Which immersive video has better quality (e.g., less noticeable distortion, stitches or blurring)?  & [VideoA, VideoB, Same, I am not sure] & VideoB & \checkmark & 45\% \\
\hline

19 & T4 & Which immersive video has better quality (e.g., less noticeable distortion, stitches or blurring)?  & [VideoA, VideoB, Same, I am not sure] & VideoA & \checkmark & 40\%\\
\hline

20 & T4 & Were the two immersive videos recorded around the same time? & [Yes, No, I am not sure] & Yes & & 75\% \\
\hline

\rev{Tutorial1} & T1 & In which immersive video are more donuts taken directly from the left three plates (indicated by the arrow)? & [VideoA, VideoB, Same, I am not sure] & VideoA &  &  \\
\hline

\rev{Tutorial2} & T2 & In which immersive video does the person move a greater distance — the woman in a rose-colored dress with sunglasses in Video A, or the man in a pink shirt with sunglasses in Video B? & [VideoA, VideoB, Same, I am not sure] & VideoB &  &  \\
\hline

\rev{Tutorial3} & T3 & Does the facing direction of the man in a blue shirt and brown trousers during the conversation differ between the two immersive videos? & [Yes, No, I am not sure] & Yes &  &  \\
\hline

\rev{Tutorial4}  & T4 & Which immersive video has better quality (e.g., less noticeable distortion, stitches or blurring)?  & [VideoA, VideoB, Same, I am not sure] & VideoB & \checkmark &  \\

\bottomrule
\end{tabular}
}
\end{table*}

\begin{table*}[h]
\centering
\small
\caption{Mean accuracy (proportion correct), standard deviation (SD), and sample size ($n$) per technique.}
\label{tab:accuracy}
\begin{tabular}{l c c c}
\toprule
\textbf{Technique} & \textbf{Mean Accuracy} & \textbf{SD} & \textbf{n} \\
\midrule
SlideInVR      & 0.64 & 0.48 & 80 \\
ToggleInVR     & 0.65 & 0.48 & 80 \\
SlideIn2D      & 0.75 & 0.44 & 80 \\
ToggleIn2D     & 0.69 & 0.47 & 80 \\
SideBySideIn2D & 0.64 & 0.48 & 80 \\
\bottomrule
\end{tabular}
\end{table*}

\begin{table*}[h]
\centering
\small
\caption{Friedman test results for perceived helpfulness of techniques across the four task types.}
\label{tab:helpfulness_stats}
\begin{tabular}{l c}
\toprule
\textbf{Task} & \textbf{$p$-value} \\
\midrule
{\,[T1]} Identify differences in the temporal occurrences of frames and ROIs.  & $0.017^{*}$ \\
{\,[T2]} Compare spatial-temporal motion of ROIs.  & 0.194 \\
{\,[T3]} Find visual differences in the visual attributes of ROIs, such as size, color and static posture.  & 0.519 \\
{\,[T4]} Identify the visual difference between the full immersive video frames.  & 0.433 \\
\bottomrule
\end{tabular}
\end{table*}

\begin{table*}[h]
\centering
\small
\caption{Friedman test results for NASA-TLX subscales and average ratings across five comparison techniques.}
\label{tab:tlx_subscales}
\begin{tabular}{l c}
\toprule
\textbf{Subscale} & \textbf{$p$-value} \\
\midrule
Mental demand     & 0.948 \\
Temporal demand   & 0.917 \\
Performance       & 0.925 \\
Effort            & 0.365 \\
Frustration       & 0.935 \\
Physical demand   & $< 0.001^{***}$ \\
\midrule
Overall workload (average) & 0.381 \\
\bottomrule
\end{tabular}
\end{table*}

\begin{table*}[h]
\centering
\small
\caption{Mean UMUX-LITE scores for each technique and pairwise Wilcoxon signed-rank test $p$-values (continuity correction). Higher scores indicate better perceived usability.}
\label{tab:umux_combined}
\begin{tabular}{l c c c c c}
\toprule
\textbf{Technique} & \textbf{Mean} & \textbf{SideBySideIn2D} & \textbf{SlideIn2D} & \textbf{SlideInVR} & \textbf{ToggleIn2D} \\
\midrule
SideBySideIn2D & 76.53 & -     & - & - & - \\
SlideIn2D      & 75.98 & 1.000 & -     & - & - \\
SlideInVR      & 69.48 & 1.000 & 1.000 & -     & - \\
ToggleIn2D     & 66.23 & 0.038$^{*}$ & 0.419 & 1.000 & - \\
ToggleInVR     & 72.19 & 1.000 & 1.000 & 1.000 & 0.604 \\
\bottomrule
\end{tabular}
\end{table*}

\begin{figure*}[hb]
    \includegraphics[width=1\textwidth]{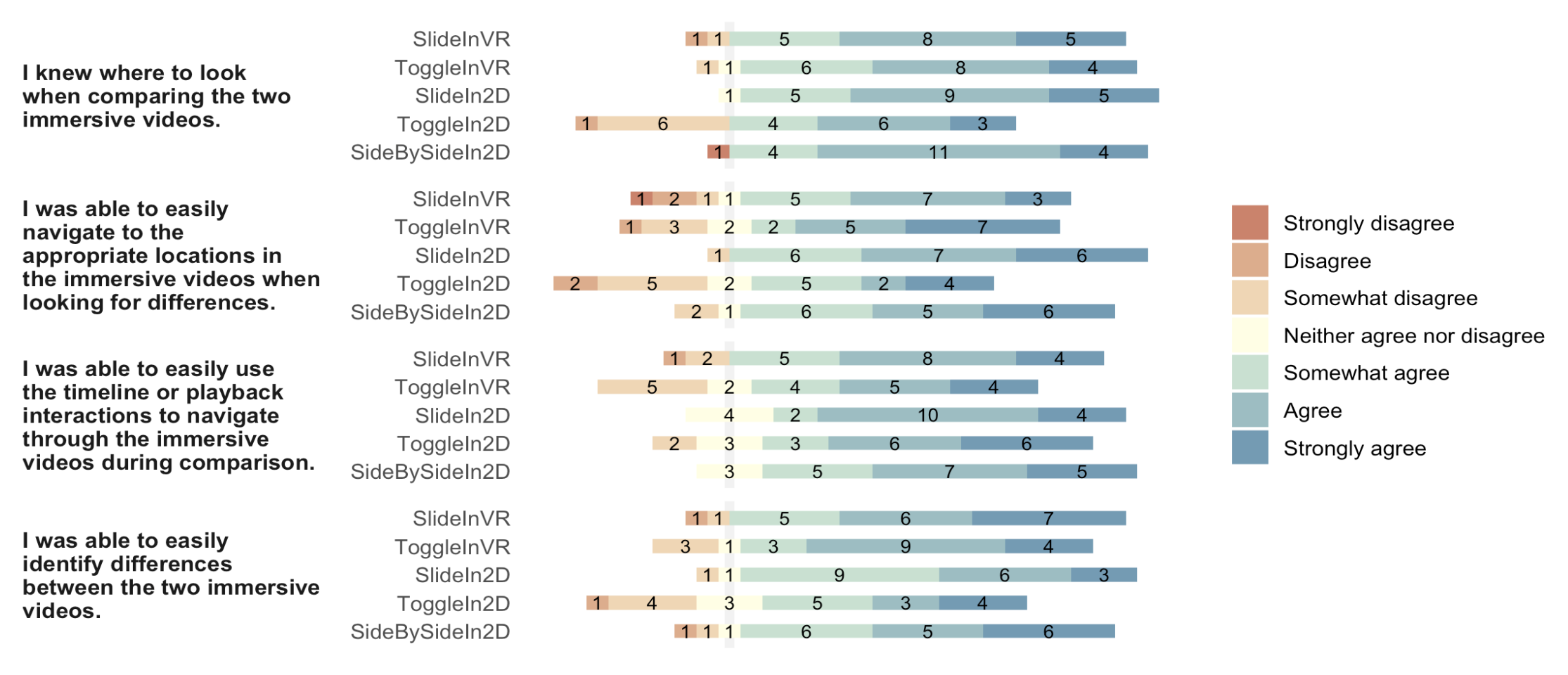}
    \vspace{-6mm}
    \caption{Stacked bar charts of how participants perceived each techniques' physical and temporal navigation experience and overall effectiveness of comparing visual difference.}
    \centering
    \label{fig:7-likert-others}
    \Description{Stacked bar charts of how participants perceived each techniques' physical and temporal navigation experience and overall effectiveness of comparing visual difference.}
    \vspace{-4mm}
\end{figure*}

\end{document}